\documentclass[12pt]{article}
\pdfoutput=1
\usepackage{jheppub_X}

\usepackage[utf8]{inputenc}

\usepackage{cleveref}
\crefname{table}{Table}{Tables}
\crefname{equation}{Eq.}{Eqs.}
\crefname{appendix}{App.}{Apps.}
\crefname{section}{Sec.}{Secs.}
\crefname{figure}{Fig.}{Figs.}
\usepackage{amsmath}
\usepackage{setspace}
\usepackage{tikz}
\usetikzlibrary{decorations.pathmorphing}
\usetikzlibrary{shapes.arrows}
\tikzset{wavy/.style={decorate, decoration=snake}}
\usepackage{tikz-feynman}
\usetikzlibrary{decorations.markings}
\usepackage[many]{tcolorbox}
\usepackage{bm}
\usepackage{xfrac}
\usepackage{pbox}
\usepackage{comment}
\usepackage{slashed}
\usepackage{physics}
\usepackage{enumitem}
\usepackage[normalem]{ulem}

\makeatletter
\g@addto@macro\bfseries{\boldmath}
\makeatother

\definecolor{colorTC}{rgb}{.2,.7,.2}

\def\eg{\textit{e.g.}}
\def\ie{\textit{i.e.}}

\newcommand{\D}{\Delta_\text{crit}}

\newcommand{\lag}{\ensuremath{\mathcal{L}}}
\newcommand{\shat}{\ensuremath{\hat{s}}}
\newcommand{\that}{\ensuremath{\hat{t}}}

\newcommand{\BSMphi}{\ensuremath{\phi}}
\newcommand{\BSMS}{\ensuremath{S}}
\newcommand{\BSMPhi}{\ensuremath{\Phi}}
\newcommand{\phiq}{\ensuremath{\phi_q}}
\newcommand{\phiqi}{\ensuremath{\phi_{q_i}}}

\newcommand{\BSMphimass}{\ensuremath{m_\phi}}

\newcommand{\muq}{\ensuremath{\mu_q}}
\newcommand{\muphi}{\ensuremath{\mu_{\phi}}}
\newcommand{\muqphi}{\ensuremath{\mu_{q\phi}}}

\newcommand{\s}{\hspace{0.8pt}}

\title{
\Large Unitarity Bounds on Effective Field Theories at the LHC
}

\author{Timothy Cohen,}
\author{Joel Doss,}
\author{and Xiaochuan Lu}

\affiliation{Institute for Fundamental Science, Department of Physics, University of Oregon, Eugene, OR 97403, USA}

\emailAdd{tcohen@uoregon.edu}
\emailAdd{jdoss@uoregon.edu}
\emailAdd{xlu@uoregon.edu}

\abstract{
Effective Field Theory (EFT) extensions of the Standard Model are tools to compute observables $\big($\emph{e.g.}~cross sections with partonic center-of-mass energy $\sqrt{\shat}\,\big)$ as a systematically improvable expansion suppressed by a new physics scale $M$.
If one is interested in EFT predictions in the parameter space where $M<\sqrt{\shat}$, concerns of self-consistency emerge, which can manifest as a violation of perturbative partial-wave unitarity.
However, when we search for the effects of an EFT at a hadron collider with center-of-mass energy $\sqrt{s}$ using an inclusive strategy, we typically do not have access to the event-by-event value of $\sqrt{\shat}$.
This motivates the need for a formalism that incorporates parton distribution functions into the perturbative partial-wave unitarity analysis.
Developing such a framework and initiating an exploration of its implications is the goal of this work.
Our approach opens up a potentially valid region of the EFT parameter space where $M \ll \sqrt{s}$.
We provide evidence that there exist valid EFTs in this parameter space.
The perturbative unitarity bounds are sensitive to the details of a given search, an effect we investigate by varying kinematic cuts.
}

\begin{document}
\maketitle
\flushbottom
\setcounter{page}{2}

\begin{spacing}{1.1}
\parskip=0ex

\section{Introduction}
\label{sec:Intro}

Searches for new physics at the LHC largely fall into two categories: (1) hunting for the signatures of the direct production of new particle(s) and (2) looking for the indirect imprint of new heavy physics on the final state distributions of Standard Model (SM) particles.  In order to parameterize the space of possible new physics effects, it is typical to utilize the theoretical frameworks of (1) Simplified Models and (2) Effective Field Theory (EFT) extensions of the SM (perhaps including other light states, \eg\ a dark matter candidate).  These theory platforms provide a principled way to design signal regions, in that they allow us to optimize sensitivity to Beyond the SM (BSM) physics.  The theory also gives an interpretation of either a null result or, all the better, a discovery of something new.  Therefore, it is of paramount importance that we ensure that our signal model frameworks are robust.  While this is typically straightforward for (perturbative) Simplified Models, it can be significantly more subtle when it comes to EFTs.

The reason that EFT validity is an interesting question stems from the starting assumption of the EFT approach:~it is necessarily a low energy approximation of a BSM theory that is associated with a dimensionful scale $M$ (the mass of a heavy BSM state in simple UV completions).  It is often the case that a search for EFT effects at the LHC yields a limit on this scale
$M\ge M_\text{limit}$ with $M_\text{limit}  < \sqrt{s}$~\cite{ATLAS:2014kci, ATLAS:2015qlt, ATLAS:2015wpv, ATLAS:2015ptl, CMS:2015rjz, CMS:2016gox, ATLAS:2016zxj, ATLAS:2017nga}, where $\sqrt{s}$ is the center-of-mass energy of the collisions.  When confronted with such a result, one should worry that the EFT approach is inconsistent (see \eg\ \cite{Kalinowski:2018oxd, Kozow:2019txg, Chaudhary:2019aim, Lang:2021hnd}).
In this work, we will investigate this question by assessing the impact of the proton's structure on one of the necessary conditions for EFT validity, namely that its scattering amplitudes satisfy perturbative partial-wave unitarity.\footnote{We emphasize that when this condition is not satisfied, what actually breaks down is the perturbative calculation itself, since we expect that the theory is fundamentally unitary. The bounds derived here should be interpreted as a necessary condition for the EFT to potentially have a perturbative UV completion.} We will provide a formalism for convolving matrix elements with Parton Distribution Functions (PDFs), and will investigate the consequences of including PDFs on the region of EFT parameter space with $M < \sqrt{s}$.

In order to better understand why PDFs are important, it is useful to recall that an EFT is an expansion that is organized using a ``power counting'' parameter $\sim 1/M$; see~\cref{subsec:PowerCounting} for a more detailed discussion.  Dimensional analysis implies that (tree-level) EFT observables yield a power series in $E/M$, where $E$ is a characteristic scale of the collision.  If $E = \sqrt{s}$ (\eg\ as it would for an $e^+ e^-$ collider) and $M \lesssim \sqrt{s}$, the series would diverge. We interpret this theoretical inconsistency as telling us that the EFT in this region of parameter space does not provide a useful description for interpreting the results of the experimental search.

At a hadron collider, the relevant scale is $E = \sqrt{\hat{s}}$, the \emph{partonic} center-of-mass energy, which varies from collision to collision as determined by the PDFs.  When designing a signal region, one is typically interested in keeping statistical fluctuations under control, which requires choosing cuts that accept events with a non-trivial range of $\sqrt{\shat}$ values. For the ensemble of events isolated by these cuts, the relevant scale on average is $E=\sqrt{\shat_\text{ave}}$, which can be much smaller than $\sqrt{s}$ due to the PDF suppression of high-energy partons.  This is why it requires a detailed investigation of a given search to determine if accounting for PDFs could salvage the EFT parameter space where $M \ll \sqrt{s}$, such that a meaningful bound can be extracted.

We emphasize that the necessity to convolve the parton-level amplitudes with the PDFs is a consequence of the following statements:
\begin{itemize}
\item The validity of an EFT depends on the experiment being performed (for our purposes here, specific to a single search region). The same EFT could be valid for one experiment, but not another, as they may use different cuts (or be at colliders with different collision energies).
\item The only physical scattering matrix elements at a hadron collider have hadrons in the initial state (not partons).
\item If the cuts used to design an experimental search region allow for a range of parton level energy scales, then one should take an ensemble average over the parton-level scattering matrix elements to determine the EFT validity for a proton level matrix element.\footnote{One might be able to isolate the ``parton'' level matrix element to a very good approximation by changing the cuts appropriately, but this would correspond to a different ``experiment.''}
\end{itemize}

In order to provide a quantitative discussion that is conceptually straightforward to interpret, we will be working with simple example UV toy models throughout this paper, leaving a detailed analysis of more realistic situations to future work.  This will allow us to derive a concrete EFT expansion by matching to the UV model, which we can use to probe the physics associated with the regions that are deemed invalid.  For an experimental search that includes a range of $\sqrt{\shat}$, PDF effects can significantly shift the perturbative unitarity bounds on EFT validity into the region where $M \ll \sqrt{s}$.  Interestingly, this conclusion begins to break down as one includes higher-and-higher terms in the $E/M$ expansion; eventually there are enough $\sqrt{\hat{s}}$ factors in the numerator to beat the strong PDF suppression at large momentum fraction.  Therefore, one of our main results is that any claim of EFT validity for a given search region requires knowing both the scale $M$ and the maximum dimension $\Delta$ (the truncation dimension) of the EFT operators.  We conclude that the question of when one can consistently use EFTs to perform searches at a hadron collider depends on both \emph{theoretical and experimental} considerations.

The rest of this paper is organized as follows. Given the extensive literature on the subject of EFT validity, we will put our work in context in \cref{sec:PrevWork}.  In \cref{sec:Models}, we discuss tree-level toy model UV completions of the benchmark pair production in \cref{eqn:pairproduction}, which are characterized by the exchange a heavy BSM scalar of mass $M$ in the $s\text{-channel}$ or $t\text{-channel}$. For each case, we match to the corresponding EFT descriptions. In \cref{sec:Unitarity}, we study perturbative partial-wave unitarity and show how to incorporate PDFs into this necessary test of EFT validity. We show that low-order EFTs can still be free of perturbative unitarity violation, even when the mass $M$ of the new physics state being integrated out is significantly below the hadronic collision energy. In \cref{sec:Xsections}, we compare the pair production cross sections predicted by the EFTs against the predictions of the UV theories. This will provide us with a way to quantitatively understand the implications of the perturbative unitarity bound that are appropriate for hadronic initial states.  In \cref{sec:Energycuts}, we vary the PDF integration limits, which allows us to explore the impact on our results as the search design becomes less inclusive. In \cref{sec:Discussions}, we conclude and discuss many future directions.  Finally, \cref{appsec:BSMpsimassLow} provides some results for the parameter space with a smaller mass for the final state particles.

\section{Strategies for Assessing EFT Validity}
\label{sec:PrevWork}

In this section, we will briefly discuss the extensive related literature, which will allow us to put the present work in context.
The subject of EFT validity is as old as the idea itself.  As the framework was being developed and its renormalization properties were being understood, \eg\ in the context of condensed matter systems~\cite{Wilson:1971bg, Wilson:1971dh, Wilson:1973jj} and gauge theories~\cite{Weinberg:1980wa}, it was always appreciated that the EFT was only meant to be applied in a limited low energy regime.  This question took on a renewed urgency in the modern era, as EFTs were being utilized as a way to design searches for new physics at the LHC, \eg\ in the context of directly producing dark matter~\cite{ATLAS:2014kci, ATLAS:2015qlt, ATLAS:2015wpv, ATLAS:2015ptl, CMS:2015rjz, CMS:2016gox, ATLAS:2016zxj, ATLAS:2017nga, Goodman:2010ku, Fox:2011pm, Fox:2012ee, Haisch:2012kf, Fox:2012ru, Papucci:2014iwa, Endo:2014mja, Garny:2015wea, Bell:2016obu, Bruggisser:2016nzw, Belyaev:2016pxe, Bishara:2016hek, Banerjee:2017wxi, Bertuzzo:2017lwt, Belwal:2017nkw, Belyaev:2018pqr, Bishara:2018vix, Trojanowski:2020xza, Fortuna:2020wwx, GAMBIT:2021rlp, Yang:2021pcf, Barman:2021hhg}, or looking for the imprint of the Standard Model EFT (SMEFT) itself~\cite{Gounaris:1994cm, Aguilar-Saavedra:2010uur, Contino:2013gna, Corbett:2014ora, Pomarol:2014dya, Azatov:2015oxa, Falkowski:2015fla, Brehmer:2015rna, LHCHiggsCrossSectionWorkingGroup:2016ypw, DiLuzio:2016sur, Contino:2016jqw, Falkowski:2016cxu, Faroughy:2016osc, Farina:2016rws, ATLAS:2017eqx, Greljo:2017vvb, Corbett:2017qgl, Alioli:2017jdo, Azatov:2017kzw, Panico:2017frx, Franceschini:2017xkh, Jin:2017guz, Alioli:2017nzr, Anders:2018oin, Aguilar-Saavedra:2018ksv, Ellis:2018gqa, CMS:2018ucw, Banerjee:2018bio, Hays:2018zze, Gomez-Ambrosio:2018pnl, Chala:2018agk, Grojean:2018dqj, Farina:2018lqo, Dawson:2018dxp, CMS:2019efc, Azatov:2019xxn, Cepeda:2019klc, Chang:2019vez, deBlas:2019rxi, Baglio:2019uty, Torre:2020aiz, CMS:2020gtj, Abu-Ajamieh:2020yqi, CMS:2020lrr, Ethier:2021ydt, CMS:2021foa, ATLAS:2021jgw, Panico:2021vav, Yang:2021gge, CMS:2021aly, Bellan:2021dcy}.  Many of these analyses explored parameter space with $M < \sqrt{s}$, prompting a variety of studies to assess the validity of the EFT description and to propose modifications to make it more robust~\cite{Shoemaker:2011vi, Busoni:2013lha, Buchmueller:2013dya, Busoni:2014sya, Busoni:2014haa, Biekotter:2014gup, Englert:2014cva, Racco:2015dxa, Gorbahn:2015gxa, Bauer:2016pug, Pobbe:2017wrj, Garcia-Garcia:2019oig, Boos:2020kqq}.  Conversely, many groups advocated to abandon the EFT approach all together in favor of Simplified Model descriptions that were clearly well defined~\cite{Buchmueller:2014yoa, Abdallah:2014hon, Malik:2014ggr, Harris:2014hga, Primulando:2015lfa, Chala:2015ama, Gupta:2015lfa, Arbey:2015hca, Abdallah:2015ter, Alves:2015mua, Kumar:2015wya, Choudhury:2015lha, Kahlhoefer:2015bea, Baker:2015qna, DeSimone:2016fbz, Boveia:2016mrp, Matsumoto:2016hbs, ATLAS:2016bek, Englert:2016joy, Jacques:2016dqz, Bruggisser:2016ixa, CMS:2016xus, Liew:2016oon, ATLAS:2017bfj, CMS:2017zts, Morgante:2018tiq, Bernal:2018nyv, LHCDarkMatterWorkingGroup:2018ufk, ATLAS:2019wdu, Ruhdorfer:2019utl, Darme:2020ral, Bertuzzo:2020rzo, Becker:2021sfd, Darme:2021xxu}.

The issues addressed by these authors essentially stem from two concerns.  The first is that when $M < \sqrt{s}$, one would expect to be able to produce the associated mediator particle directly, since it is the mediator's mass that sets the scale $M$.  This opens up new and often more powerful ways to search for the signatures of the associated model.  We have nothing novel to say about this important effect.  However, we remind the reader that in the narrow width approximation, the production of the heavy on-shell states would not interfere with the processes captured by the EFT.  Therefore, including these additional direct mediator production processes would lead to a stronger limit, in principle, than what one would obtain by using the EFT alone.  Since the EFT limit does not require specifying a concrete UV completion, it can be applied to a broader class of models without a dedicated recasting effort. For these reasons, we advocate that EFT searches are still useful in their own right, although care must be taken with regards to their interpretation.

The second concern is of direct relevance to the study we perform here.  Recognizing that $\sqrt{\hat{s}}$ is the quantity of interest when testing for EFT validity, a variety of proposals were put forward that shared a theme of ``cutting away high energy events.''  In other words, the kinematics were restricted so as to avoid the region of phase space where the EFT validity was in question.

For example, the authors of \cite{Racco:2015dxa} proposed to cut away any event with $\sqrt{\hat{s}} > M$ (at simulation truth level) when computing the signal rates. Incorporating a maximum allowed value of $\sqrt{\hat{s}}$ within the signal simulation ensures the validity of the EFT, at the expense of reducing the EFT prediction significantly.  One could even consider the cut on maximum allowed $\sqrt{\hat{s}}$ as an additional parameter of the signal model. One could vary this additional parameter to see how sensitive a given search is to such high energy events.\footnote{We thank Markus Luty for emphasizing this point of view to us.}  Although the limits derived with this approach are strictly valid, the resulting bounds on $M$ can be artificially conservative.

Another group~\cite{Busoni:2013lha, Busoni:2014sya, Busoni:2014haa}, proposed to cut on the observable kinematics of the final state as a proxy for removing high energy events.  In the same spirit, experimentalists have applied a high energy cut parameter to some of their EFT analyses, investigating the robustness of the limits they derive when this parameter is varied~\cite{ATLAS:2014kci, ATLAS:2015qlt, ATLAS:2015wpv, ATLAS:2015ptl, CMS:2015rjz, CMS:2016gox, ATLAS:2016zxj, ATLAS:2017nga, Abercrombie:2015wmb}. An alternative strategy to ``unitarize'' the EFT has also been employed in some experimental searches~\cite{ATLAS:2016snd, CMS:2020gfh}.  Note that essentially all previous validity studies are truncated at the leading order EFT dimension; see~\cite{Biekotter:2014gup} for a notable exception.

Our focus here is on exploring the impact of PDFs on the perturbative partial-wave unitarity bound.  We emphasize that while perturbative partial-wave unitarity provides a good proxy for the question of EFT validity, satisfying this condition is necessary but not sufficient. The key insight of this paper is to leverage the fact that a typical signal region includes events with a range of associated $\sqrt{\hat{s}}$.  Therefore, one must incorporate an ensemble of events when diagnosing EFT validity --- when working at a hadron collider, this can be accounted for by properly treating PDF effects.  We will show that (for sufficiently inclusive signal regions) perturbative unitarity bounds on EFTs are typically insensitive to cutting away high energy events (when the cut is applied to both signal and background), which we take to be a sign that the EFT validity is being saved by the PDF suppression at high momentum fraction.  One of the goals of this work is to make this intuition precise.

\section{Benchmark Process and Toy Models}
\label{sec:Models}

As we emphasized above, our goal in this paper is to study the impact of having an ensemble of events with various values of $\sqrt{\hat{s}}$ on the question of EFT validity.  To this end, we will focus our attention on the benchmark pair production process
\begin{equation}
\left(\, p\,p \to \BSMphi\,\BSMphi^\dagger\, \right) \;=\; \sum_{\{q,\,\bar{q}\}\,\in\,p}\, \left(\, \phiq\,\phiq^\dagger \to \BSMphi\,\BSMphi^\dagger\, \right) \,.
\label{eqn:pairproduction}
\end{equation}
For simplicity, we will study this question using two simple toy model UV theories.  At tree level, these models are characterized by how they generate the benchmark pair production by either the $t$-channel or $s$-channel exchange of a heavy BSM scalar, as illustrated in \cref{fig:tchannelmodel,fig:schannelmodel}, respectively.  We will then match these theories onto the subset of EFT operators that contribute to \cref{eqn:pairproduction} at tree-level.

Note that to minimize the technical aspects of what follows, we have chosen to work with scalar ``quarks'' $\phiq, \phiq^\dagger$ in the initial state (not to be confused with ``squarks'' in supersymmetric theories).  Specifically, we will be using the $q, \bar{q}$ quark PDFs when investigating the interpretation of the EFT parameter space.  This has the benefit that the analytic formulas will be very simple, at the obvious expense of not being fully realistic.\footnote{We will perform the analysis for fermionic initial and final states in a future paper.  We also anticipate that we will find similar conclusions if the production process is dominated by a gluon or mixed quark/gluon initial state, which we also plan to study in a future paper.}

We will make an additional simplifying choice in what follows.  When we integrate out a heavy state, the leading order contribution to the EFT Lagrangian appears at dimension 4, since we are working with pure scalar toy theories.  Note that the operators of interest here are those that lead to cross section growth, which have dimension $> 4$.  Therefore, we will tune a Lagrangian quartic parameter against the EFT contribution so that the leading contribution to the 2-to-2 scattering of interest comes from a dimension 6 operator.  This makes our results much more intuitive, and also more relevant to the realistic case where the quarks (and perhaps also the final state particles) are fermions.

The final state $\phi$ is a (relatively light) BSM scalar.  It could be a dark matter candidate, some other BSM state, or even an SM particle (as it would be in the case of SMEFT searches); all that matters in what follows is that it is a scalar, and otherwise we are agnostic about its identity. The only EFT operators that contribute to the benchmark pair production process in \cref{eqn:pairproduction} are those involving extra derivatives with a fixed number of fields (leading to a $1/M$ expansion).  We will briefly comment on the relation to the EFT operators that involve more powers of fields (leading to a $1/\Lambda$ expansion) in \cref{subsec:PowerCounting}.

\subsection{$t\s$-Channel Model}
\label{subsec:tModels}

A model that results in $t$-channel pair production utilizes a heavy complex scalar mediator $\BSMPhi$ that couples to the scalar quarks $\phi_q$ and the BSM singlet scalar $\BSMphi$ through a tri-linear interaction:
\begin{align}
\lag_{t,\s\text{UV}} &\supset \lag_\text{SM} - \BSMphi^\dagger\! \left( \partial^2 +\BSMphimass^2 \right)\! \BSMphi -\lambda_{q\phi}\! \left( \phiq^\dagger\BSMphi \right)\!\! \left( \BSMphi^\dagger\phiq \right) \notag\\[5pt]
&\quad - \BSMPhi^\dagger\!\left( D^2 + M^2 \right)\!\BSMPhi - \muqphi\s\phiq^\dagger\BSMphi\,\BSMPhi - \muqphi^*\s\BSMPhi^\dagger\,\BSMphi^\dagger \phiq \,,
\label{eqn:lagtUV}
\end{align}
where $\lag_\text{SM}$ is the SM Lagrangian (including the kinetic term for the scalar quarks), $D_\mu$ is a gauge covariant derivative, $\BSMphimass$ is the mass of the BSM singlet scalars, $\lambda_{q\phi}$ is a cross quartic coupling, $M$ is the mass of the heavy scalar mediator $\BSMPhi$, and $\muqphi$ is a tri-linear coupling.  Since $\BSMphi$ is a SM singlet, the heavy complex scalar $\BSMPhi$ needs to have the same SM charge as the scalar quark $\phi_q$ to ensure that the tri-linear coupling is gauge invariant. For concreteness, we will assume a universal coupling to the $\phi_q$ with $q \in \{d_R, s_R, b_R\}$. This choice has a minimal impact on our conclusions.

As depicted in \cref{fig:tchannelmodel}, the EFT description for the $t$-channel pair production process can be obtained by expanding the propagator:
\begin{equation}
\frac{1}{\that-M^2} \to -\frac{1}{M^2}\sum_{r=0}^k \left(\frac{\that}{M^2}\right)^r \,,
\end{equation}
where $k$ corresponds to the desired EFT truncation order. Using the 2-to-2 kinematic constraints, we have
\begin{equation}
\that = \left(p_1-p_3\right)^2 = \left(p_2-p_4\right)^2 \,,
\end{equation}
which implies that the relevant part of the EFT Lagrangian is given by
\begin{align}
\lag_{t,\,\text{EFT}} &\supset \lag_\text{SM} - \BSMphi^\dagger\! \left(\partial^2 + \BSMphimass^2 \right)\! \BSMphi -\lambda_{q\phi}\! \left( \phiq^\dagger\BSMphi \right)\!\! \left( \BSMphi^\dagger\phiq \right) + \frac{\left|\muqphi\right|^2}{M^2} \sum_{r=0}^{k+1} \big(\phiq^\dagger\BSMphi\big)\! \left( -\frac{\partial^2}{M^2} \right)^r\! \big(\BSMphi^\dagger \phiq\big) \notag\\[5pt]
&= \lag_\text{SM} - \BSMphi^\dagger\! \left(\partial^2 + \BSMphimass^2 \right)\! \BSMphi - \frac{\lambda_{q\phi}}{M^2} \sum_{r=0}^k \big(\phiq^\dagger\BSMphi\big)\! \left( -\frac{\partial^2}{M^2} \right)^r\! \partial^2 \big(\BSMphi^\dagger \phiq\big) \,.
\label{eqn:lagtEFT}
\end{align}
In the second line, we have set
\begin{align}
\lambda_{q\phi} = \frac{\left|\muqphi\right|^2}{M^2}\,,
\label{eq:lambdaQPhi}
\end{align}
in order to tune away the dimension-4 contribution, and have relabeled the summation index $r$.\footnote{Note that the $\partial^2$ in this EFT Lagrangian should technically be promoted to $D^2$ to form gauge-invariant effective operators. However, the extra terms that result contain additional gauge bosons and hence do not contribute to $\phiq\,\phiq^\dag \to \BSMphi\,\BSMphi^\dag$ at tree level, and so we do not include them here.}

\begin{figure}[t]
\centering
\begin{tikzpicture}[baseline=($0.5*(i1)+0.5*(i2)$)]
\begin{feynman}
\vertex(a);
\vertex[below = 1.25cm of a](b);
\vertex[above left = 1.15cm of a](i1){$\phi_q$};
\vertex[below left = 1.15cm of b](i2){$\phi_q^\dag$};
\vertex[above right = 1.15cm of a](f1){$\phi^\dag$};
\vertex[below right = 1.15cm of b](f2){$\phi$};
\diagram*{
(i1)--[charged scalar](a)--[charged scalar](f1),
(a)--[charged scalar, edge label=$\;\Phi$](b),
(f2)--[charged scalar](b)--[charged scalar](i2)};
\end{feynman}
\end{tikzpicture}
\hspace{1cm}
\begin{tikzpicture}[baseline=0cm]
\node [single arrow, draw, fill=gray, minimum height=2cm] {};
\end{tikzpicture}
\hspace{1cm}
\begin{tikzpicture}[baseline=($0.5*(i1)+0.5*(i2)$)]
\begin{feynman}
\vertex[dot](a){};
\vertex[above left = 1.7cm of a](i1){$\phi_q$};
\vertex[below left = 1.7cm of a](i2){$\phi_q^\dag$};
\vertex[above right = 1.7cm of a](f1){$\phi^\dag$};
\vertex[below right = 1.7cm of a](f2){$\phi$};
\diagram*{
(i1)--[charged scalar](a)--[charged scalar](i2),
(f2)--[charged scalar](a)--[charged scalar](f1)};
\end{feynman}
\end{tikzpicture}
\caption{Pair production $\phi_q \phi_q^\dag \to\phi\s\phi^\dag$ through the $t$-channel exchange of a heavy complex scalar $\BSMPhi$. The EFT description can be obtained by expanding and truncating the $t$-channel propagator, yielding a series of local operators.}
\label{fig:tchannelmodel}
\end{figure}

The maximum dimension of the EFT operators $\Delta$ is related to the truncation order $k$:
\begin{equation}
\Delta = 6 + 2k \,.
\label{eq:Delta6}
\end{equation}
At the lowest truncation order $k=0$, our pair production process is modeled by the dimension-six operator
\begin{equation}
\mathcal{O}_6 = -\frac{\lambda_{q\phi}}{M^2}\! \left(\phiq^\dagger \BSMphi\right)\! \partial^2\! \left(\BSMphi^\dagger \phiq\right) = -\frac{1}{\Lambda^2} \left(\phiq^\dagger \BSMphi\right)\! \partial^2 \!\left(\BSMphi^\dagger \phiq\right) \,.
\label{eqn:LambdaDef}
\end{equation}
The contribution from EFT operators are often said to be characterized by the scale $\Lambda \equiv M/\sqrt{\lambda_{q\phi}}$, which could be much higher than the mediator mass $M$ in the weakly coupled limit $\lambda_{q\phi} \ll 1$. However, note that the contribution from higher orders in the EFT expansion for the 2-to-2 process of interest here is actually controlled by the suppression factor $E^2/M^2$ as opposed to $E^2/\Lambda^2$, see~\cref{eqn:lagtEFT}. This distinction is important for interpreting analyses that go beyond dimension-6, see the discussion in~\cref{subsec:PowerCounting} below.

\subsection{$s\s$-Channel Model}
\label{subsec:sModels}

Next, we can write down a model that will yield $s$-channel production of a pair of BSM singlet scalars $\phi$.  This can be accomplished by introducing a heavy real singlet scalar mediator $\BSMS$ that couples to the scalar quarks $\phi_{q_i}^\dag \phi_{q_i}$ and to $\phi^\dag \phi$:
\begin{align}
\lag_{s,\,\text{UV}} &\supset \lag_\text{SM} - \BSMphi^\dagger\! \left(\partial^2 +\BSMphimass^2 \right)\! \BSMphi - \lambda_q\! \left( \phi_{q_i}^\dag \phi_{q_i} \right)\!\! \left( \BSMphi^\dagger \BSMphi \right) \notag\\[5pt]
&\quad - \frac12\BSMS\! \left(\partial^2 + M^2\right)\! \BSMS - \muq\s\phi_{q_i}^\dag \phi_{q_i}\BSMS - \muphi\s \BSMphi^\dagger \BSMphi\s\BSMS \,,
\label{eqn:lagsUV}
\end{align}
where $\lag_\text{SM}$ is the SM Lagrangian, $m_\phi$ is the mass of the BSM singlet scalars, $\lambda_q$ is a cross quartic coupling, $M$ is the mass of the heavy scalar mediator $\BSMS$, $\muq$ and $\muphi$ are tri-linear couplings, and we interpret $\phi_{q_i}^\dag \phi_{q_i}$ as the sum over all species and flavors of quarks in the SM.  Note that we have only included the leading interactions that are relevant for our purposes here, see \cref{subsec:PowerCounting} below for a related discussion.

As depicted in \cref{fig:schannelmodel}, the EFT description for the tree-level $s$-channel pair production can be obtained by expanding and truncating the propagator:
\begin{equation}
\frac{1}{\shat-M^2} \to -\frac{1}{M^2}\sum_{r=0}^k \left(\frac{\shat}{M^2}\right)^r \,,
\end{equation}
where we are introducing the parameter $k$ as in \cref{subsec:tModels} above, and we have set the width of $S$ to zero. Since
\begin{equation}
\shat = \left(p_1+p_2\right)^2 = \left(p_3+p_4\right)^2 \,,
\end{equation}
\clearpage
\noindent for 2-to-2 kinematics, we infer that the EFT Lagrangian is given by
\begin{align}
\lag_{s,\s\text{EFT}} &\supset \lag_\text{SM} - \BSMphi^\dagger\! \left( \partial^2 +\BSMphimass^2 \right)\! \BSMphi - \lambda_q\! \left( \phi_{q_i}^\dag \phi_{q_i} \right)\!\! \left( \BSMphi^\dagger \BSMphi \right) + \frac{\muq\s\muphi}{M^2}\sum_{r=0}^{k+1}\! \left(\phiqi^\dagger \phiqi\right)\!\! \left(-\frac{\partial^2}{M^2} \right)^r\!\! \big(\BSMphi^\dagger \BSMphi\big) \notag\\[5pt]
&= \lag_\text{SM} - \BSMphi^\dagger\! \left(\partial^2 +\BSMphimass^2 \right)\! \BSMphi - \frac{\lambda_q}{M^2}\sum_{r=0}^{k}\!\left(\phiqi^\dagger \phiqi\right)\!\! \left(-\frac{\partial^2}{M^2} \right)^r\!\partial^2\big(\BSMphi^\dagger \BSMphi\big) \,.
\label{eqn:lagsEFT}
\end{align}
In the second line, we have again tuned the quartic coupling
\begin{align}
\lambda_q = \frac{\muq\s\muphi}{M^2}\,,
\label{eq:lambdaqTune}
\end{align}
to cancel the dimension-4 effect. Of course the EFT generates many additional operators beyond the ones written here.  However, none of these contribute to the pair production $\phiq\s \phiq^\dag \to \phi\s\phi^\dag$ at tree level, and so we do not write them explicitly.  Similar to the $t$-channel model, we can identify a scale $\Lambda = M/\sqrt{\lambda_q}$, which sets the overall rate.  Once that is specified, the EFT operators relevant here are controlled by a $1/M$ expansion.

\begin{figure}[t!]
\centering
\begin{tikzpicture}[baseline=($0.5*(i1)+0.5*(i2)$)]
\begin{feynman}
\vertex(a);
\vertex[right = 1.5cm of a](b);
\vertex[above left = 1.25cm of a](i1){$\phi_q$};
\vertex[below left = 1.25cm of a](i2){$\phi^\dag_q$};
\vertex[above right = 1.25cm of b](f1){$\phi^\dag$};
\vertex[below right = 1.25cm of b](f2){$\phi$};
\diagram*{
(i1)--[charged scalar](a)--[charged scalar](i2),
(a)--[scalar, edge label'=$S$](b),
(f2)--[charged scalar](b)--[charged scalar](f1)};
\end{feynman}
\end{tikzpicture}
\hspace{1cm}
\begin{tikzpicture}[baseline=0cm]
\node [single arrow, draw, fill=gray, minimum height=2cm] {};
\end{tikzpicture}
\hspace{1cm}
\begin{tikzpicture}[baseline=($0.5*(i1)+0.5*(i2)$)]
\begin{feynman}
\vertex[dot](a){};
\vertex[above left = 1.7cm of a](i1){$\phi_q$};
\vertex[below left = 1.7cm of a](i2){$\phi^\dag_q$};
\vertex[above right = 1.7cm of a](f1){$\phi^\dag$};
\vertex[below right = 1.7cm of a](f2){$\phi$};
\diagram*{
(i1)--[charged scalar](a)--[charged scalar](i2),
(f2)--[charged scalar](a)--[charged scalar](f1)};
\end{feynman}
\end{tikzpicture}
\caption{Pair production $\phi_q \phi_q^\dag \to\phi\s\phi^\dag$ through the $s$-channel exchange of a heavy singlet scalar $\BSMS$. The EFT description can be obtained by expanding and truncating the $s$-channel propagator, yielding a series of local operators. }
\label{fig:schannelmodel}
\end{figure}

\subsection{On the $1/M$ Versus $1/\Lambda$ EFT Expansions}
\label{subsec:PowerCounting}

As explained in \cref{sec:Intro}, current limits on $M$ derived by LHC searches are typically around a few TeV, well below the collider energy, thereby raising the question of EFT validity for practical situations.  Using the toy models discussed in \cref{subsec:sModels,subsec:tModels}, which characterize the corrections to our benchmark pair production process in~\cref{eqn:pairproduction}, we will be focused on the effects of the $E/M$ power series in the EFT expansions (see \cref{eqn:lagsEFT,eqn:lagtEFT}). One may be concerned that our results that follow are a special feature of this specific choice. In fact, a different but very typical expectation for a general EFT expansion is that effects from higher dimension operators would come with powers of $E/\Lambda$ (instead of $E/M$), where $\Lambda \equiv M/\sqrt{\lambda}$ characterizes the effects of dimension six (leading order) operators with $\lambda = \lambda_{q\phi}$ or $\lambda_q$ for the $t$- and $s$-channel models respectively. Since it is possible that $\Lambda \gg M$ in the weakly coupled limit $\lambda \ll 1$, it could also be the case that $\Lambda > \sqrt{s}$.  Therefore, this expectation might make one wonder if EFT validity is actually a problem. In this subsection, we address this potential concern.

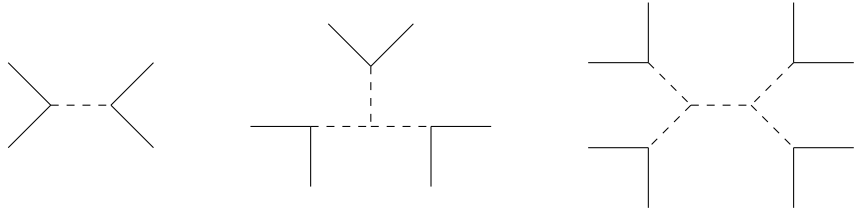
\begin{figure}[t!]
\centering
\begin{tikzpicture}[baseline=($0.5*(i1)+0.5*(i2)$)]
\begin{feynman}
\vertex(a);
\vertex[right = 0.8cm of a](b);
\vertex[above left = 0.8cm of a](i1);
\vertex[below left = 0.8cm of a](i2);
\vertex[above right = 0.8cm of b](f1);
\vertex[below right = 0.8cm of b](f2);
\diagram*{
(i1)--(a)--(i2),
(a)--[scalar](b),
(f2)--(b)--(f1)};
\end{feynman}
\end{tikzpicture}
\hspace{1cm}
\begin{tikzpicture}[baseline=($0.5*(f1)+0.5*(f4)$)]
\begin{feynman}
\vertex(a);
\vertex[right = 0.8cm of a](b);
\vertex[above = 0.8cm of b](c);
\vertex[right = 0.8cm of b](d);
\vertex[left = 0.8cm of a](i1);
\vertex[below = 0.8cm of a](i2);
\vertex[above right = 0.8cm of c](f1);
\vertex[above left = 0.8cm of c](f2);
\vertex[right = 0.8cm of d](f3);
\vertex[below = 0.8cm of d](f4);
\diagram*{
(i1)--(a)--(i2),
(a)--[scalar](b),
(c)--[scalar](b)--[scalar](d),
(f2)--(c)--(f1),
(f4)--(d)--(f3)};
\end{feynman}
\end{tikzpicture}
\hspace{1cm}
\begin{tikzpicture}[baseline=(b)]
\begin{feynman}
\vertex(a1);
\vertex[below right = 0.8cm of a1](b);
\vertex[below left = 0.8cm of b](a2);
\vertex[right = 0.8cm of b](c);
\vertex[above right = 0.8cm of c](d1);
\vertex[below right = 0.8cm of c](d2);
\vertex[above = 0.8cm of a1](i1);
\vertex[left = 0.8cm of a1](i2);
\vertex[left = 0.8cm of a2](i3);
\vertex[below = 0.8cm of a2](i4);
\vertex[above = 0.8cm of d1](f1);
\vertex[right = 0.8cm of d1](f2);
\vertex[right = 0.8cm of d2](f3);
\vertex[below = 0.8cm of d2](f4);
\diagram*{
(i1)--(a1)--(i2),
(i3)--(a2)--(i4),
(a1)--[scalar](b)--[scalar](a2),
(b)--[scalar](c),
(d1)--[scalar](c)--[scalar](d2),
(f2)--(d1)--(f1),
(f4)--(d2)--(f3)};
\end{feynman}
\end{tikzpicture}
\caption{A series of EFT operators with more powers of fields generated by insertions of the self trilinear coupling in the $s$-channel model. Dashed lines denote the heavy scalar mediator $S$; solid lines denote light particles, either scalar quarks or the BSM singlet scalars $\BSMphi$.}
\label{fig:fieldexpansion}
\end{figure}

In fact, the EFT Lagrangians (\cref{eqn:lagsEFT,eqn:lagtEFT}) obtained from the toy models are somewhat special, in the sense that higher order EFT operators come with strictly more powers of \emph{derivatives}. This is not the case for a generic EFT expansion. Taking, for example, the $s$-channel toy model in \cref{eqn:lagsUV}, we can include an allowed self trilinear coupling for the heavy scalar mediator $S$
\begin{align}
\lag_{s,\,\text{UV}} \supset \frac{1}{3!}\s a\s S^{3} \,.
\end{align}
Insertions of this vertex could generate a series of EFT operators with more powers of \emph{fields} at each order, as illustrated in \cref{fig:fieldexpansion}. In general, higher order EFT operators could contain more powers of either \emph{derivatives} or \emph{fields}. Operators of the former type would obviously contribute with more powers of $E/M$.  Operators of the latter type would contribute with either more powers of $E/\Lambda$, or a mix of $E/\Lambda$ and $E/M$ factors.

To see an example of this, we consider the series of effective operators generated by the diagrams depicted in \cref{fig:fieldexpansion}. We note that these operators have different external states, and hence the associated amplitudes do not interfere with each other.  One way to compare the size of their contributions is to consider the inclusive cross section $\phiq\phiq^\dag \to \text{anything}$.  In this case, taking into account phase space, one can show that operators with more insertions of the self trilinear coupling lead to the pattern
\begin{equation}
\text{each cubic insertion}\,\, \Rightarrow\,\, \frac{\muphi\s a}{M^2}\s\frac{E^2}{M^2}\,.
\label{eq:CubicInsertion}
\end{equation}
Analogous to \cref{eqn:LambdaDef}, one could define a ``$\Lambda$'' for each UV coupling:\footnote{We note that $M$ and $\Lambda$ have different units when the factors of $\hbar$ are restored, see \eg\/~\cite{Giudice:2007fh, Contino:2016jqw}. This underscores the point we are trying to make here, namely that they control two different categories of EFT expansions.}
\begin{equation}
\frac{1}{\Lambda_{\BSMphi}} \equiv \frac{\muphi}{M^2} \,,\qquad\text{and}\qquad
\frac{1}{\Lambda_a} \equiv \frac{a}{M^2} \,.
\end{equation}
Using these $\Lambda$'s, we can rewrite \cref{eq:CubicInsertion}:
\begin{equation}
\frac{\muphi\s a}{M^2}\, \frac{E^2}{M^2} = \frac{E}{\Lambda_{\BSMphi}}\s \frac{E}{\Lambda_a} \,.
\label{eq:CubicInsertion2}
\end{equation}
We see that in this specific example, operators with more fields would contribute with more powers of $E/\Lambda$, agreeing with the typical expectation. In general cases, contributions from operators with more fields and derivatives could come with a mix of $E/\Lambda$ and $E/M$ factors.

The above discussion shows that the typical expectation that the EFT expansion is governed by $E/\Lambda$ alone is incomplete; for certain sets of operators it is true (such as in \cref{eq:CubicInsertion2}), but other operators could be governed by an $E/M$ expansion (such as in \cref{eqn:lagsEFT,eqn:lagtEFT}). Therefore, when $\Lambda \gg M$ as it is for weakly coupled scenarios, our choice to focus on EFTs for the benchmark pair production process in~\cref{eqn:pairproduction} is exploring the \emph{most dangerous} contributions to the question of EFT validity. As we will show in the rest of this paper, even when considering these most dangerous operators at the limit of perturbativity, EFTs parameter space with $M<\sqrt{s}$ can still be a valid framework for BSM searches using inclusive signal regions, provided that the EFT expansion is not extended to a ridiculously high order.

\section{Partial-Wave Unitarity Bounds}
\label{sec:Unitarity}

In this section, we investigate the impact of incorporating PDFs into perturbative partial-wave unitarity bounds.  This will allow us to explore the interplay of perturbative unitarity violation, which emerges when one probes an EFT at high energies, and PDFs, which act to suppress the production of those problematic high energy events. To this end, we will need to develop a formalism to incorporate PDFs into the partial wave perturbative unitarity test.  Specifically, we will generalize the standard partial wave perturbative unitarity argument that applies to pure initial quantum states (appropriate for parton-level scattering) to the case of mixed/ensemble initial quantum states (appropriate for hadron-level scattering).

The results presented in this section are obtained by working with the EFTs discussed in \cref{sec:Models}. Thus, everything in this section is specific to our simple toy UV completions. The approach of using perturbative unitarity violation to determine EFT validity is often viewed as a bottom-up consistency test.  It provides a necessary (but not sufficient) condition that the EFT is a well behaved quantum theory.   Although we are providing model specific results here, the conclusions we will draw are expected to apply to general EFTs.

First, \cref{subsec:PartonUnitarity} presents the perturbative unitarity constraints derived using parton-level scattering for the UV theories and the EFTs detailed in \cref{sec:Models}. Then we turn to \cref{subsec:HadronUnitarity}, where we show how to incorporate PDFs into the $s$-wave perturbative unitarity test.  In \cref{subsec:UnitarityResults}, we apply this technique to numerically explore when these constraints are violated.  We are particularly interested in scenarios where the new physics scale is below the hadron collider energy $M \ll \sqrt{s}$, and we will show that perturbative unitarity is violated when the EFT truncation order $k$ is sufficiently large. However, when $k$ is small and the signal region is sufficiently inclusive, the theory passes the perturbative partial-wave unitarity test due to the PDF suppression of high-energy partons. Therefore, such low-order EFTs are free of perturbative unitarity violation. This will allow us to derive upper bounds on the $(M, \Delta)$ parameter space, which will denote regions of parameter space where the EFT predictions can not be trusted, \ie, regions where one cannot place experimental bounds on the EFT.

\subsection{Partonic Initial State}
\label{subsec:PartonUnitarity}

Tree-level perturbative unitarity constraints on a scattering process at the parton level can be obtained by checking that the $S$-matrix is unitary. In principle, this can be done for each component in the partial wave expansion of the amplitude. In this paper, we focus on the $s$-wave component for simplicity. It is somewhat tedious to keep track of all the $S$-matrix components corresponding to the various spin configurations when fermions are involved in the scattering (see example calculations in \cite{Endo:2014mja, DiLuzio:2016sur, Bell:2016obu, Kumar:2015wya, Banta:2021dek} and \cite{Chang:2019vez,Abu-Ajamieh:2020yqi} for recent reviews). Minimizing this technical complication is the reason for studying the scalar toy models introduced in \cref{sec:Models}.  As we stated there, we will give $\phiq$($\phiq^\dagger$) the same proton PDFs as the quark fields $q$($\bar{q}$), and will treat them as massless.

\clearpage

The $s$-wave perturbative unitarity condition on the parton-level pair production in \cref{eqn:pairproduction} can be succinctly summarized as:\footnote{Our notation for the normalized $s$-wave amplitude $\mathcal{M}$ here follows that in \cite{Chang:2019vez, Abu-Ajamieh:2020yqi}, which differs from the $a_0$ notation in {\it e.g.}~\cite{Schwartz:2014sze} by a factor of two: $\mathcal{M}=2a_0$.}
\begin{subequations}
\begin{align}
\Omega(\shat) &\equiv \left|\mathcal{M}(\shat)\right|^2 \le 1 \,, \\[5pt]
\mathcal{M}(\shat) &\equiv \left(\frac{\shat-4\BSMphimass^2}{\shat}\right)^{1/4} \frac{1}{16\pi} \int_{-1}^1 \dd (\cos\theta)\mathcal{A}\!\left(\cos\theta\right) \,.
\end{align}%
\label{eqn:unitaritycondition}%
\end{subequations}%
Here $\mathcal{A}$ is the usual scattering amplitude and $\mathcal{M}$ is its $s$-wave component. In what follows, we will apply this test to the $t$-channel and $s$-channel production models detailed in \cref{sec:Models}, to determine if perturbative partial-wave unitarity is satisfied for this process.

\subsubsection*{$t\s$-Channel UV Theory}

For the $t$-channel production UV model in \cref{eqn:lagtUV}, the scattering amplitude is
\begin{equation}
\mathcal{A}_t = - \frac{\left|\muqphi\right|^2}{\that - M^2} -\lambda_{q\phi} = -\lambda_{q\phi}\s \frac{\that}{\that - M^2} \,,
\label{eqn:scalarAt}
\end{equation}
where we have applied \cref{eq:lambdaQPhi} to tune away the dimension-4 contribution.  Following \cref{eqn:unitaritycondition} and using the kinematic relation
\begin{equation}
\that = \BSMphimass^2 - \frac{\shat}{2}\! \left(1 - \sqrt{ \frac{\shat-4\BSMphimass^2}{\shat} }\, \cos\theta \right) \,,
\label{eqn:thatintheta}
\end{equation}
the $s$-wave component is given by integrating over the scattering angle:
\begin{align}
\mathcal{M}_t &= \left(\frac{\shat-4\BSMphimass^2}{\shat}\right)^{1/4} \frac{1}{16\pi} \int_{-1}^1 \dd (\cos\theta) \mathcal{A}_t \notag\\[7pt]
&= \frac{\lambda_{q\phi}}{8\pi}\s \frac{M^2}{\shat}\! \left(\frac{\shat-4\BSMphimass^2}{\shat}\right)^{-1/4} \left[ \log\frac{1+\kappa_+}{1+\kappa_-} - \left( \kappa_+ - \kappa_- \right) \right] \,,
\label{eqn:Mt}
\end{align}
where we have introduced the dimensionless quantities
\begin{equation}
\kappa_\pm \equiv \frac{\shat}{4M^2}\! \left( 1 \pm \sqrt{\frac{\shat - 4\BSMphimass^2}{\shat}} \,\right)^2 \,.
\label{eqn:auxiliary}
\end{equation}
 This leads to the parton-level $s$-wave perturbative unitarity condition for the $t$-channel UV model
\begin{equation}
\hat\Omega_{t,\s\text{UV}}\!\left(\shat\right) = \left|\mathcal{M}_t\right|^2 = \frac{\lambda_{q\phi}^2}{64\pi^2}\s \frac{M^4}{\shat^2}\!\left(\frac{\shat-4\BSMphimass^2}{\shat}\right)^{-1/2}\, \left[ \log\frac{1+\kappa_+}{1+\kappa_-} - \left( \kappa_+ - \kappa_- \right) \right]^2 \le 1 \,.
\label{eqn:hatOmegatUV}
\end{equation}
For most of the numerical results that follow, we will set
\begin{align}
\lambda_{q\phi} = 8 \pi \,,
\label{eq:tChParam}
\end{align}
which is compatible with the condition in \cref{eqn:hatOmegatUV}, see \cref{fig:OmegatUV}.

\subsubsection*{$t\s$-Channel EFT}
We can obtain the corresponding EFT result $\hat\Omega_{t,\,\text{EFT}}(\shat)$ by repeating the above calculation for the EFT Lagrangian in \cref{eqn:lagtEFT}. Equivalently, we can expand the UV result for the $s$-wave amplitude $\mathcal{M}_t$ in \cref{eqn:Mt} as a power series in $1/M^2$ up to some order $k$ (or dimension $\Delta=6+2k$).\footnote{Note that when deriving the perturbative unitarity condition, the EFT expansion should be applied to the $s$-wave amplitude $\mathcal{M}_t$, not to its modulus square $\hat\Omega_{t,\s\text{UV}}\equiv\left|\mathcal{M}_t\right|^2$. The latter should always be kept as a complete norm square for each choice of $k$.} This yields
\begin{equation}
\mathcal{M}_{t}^{[k]} = \frac{\lambda_{q\phi}}{8\pi}\, \frac{M^2}{\shat}\, \left(\frac{\shat-4\BSMphimass^2}{\shat}\right)^{-1/4}\, \sum_{r=0}^k \frac{(-1)^{r+1}}{r+2} \left(\kappa_+^{r+2} - \kappa_-^{r+2}\right) \,,
\end{equation}
which leads to the EFT $s$-wave perturbative unitarity condition
\begin{equation}
\hat\Omega_{t,\,\text{EFT}}^{[k]}\!\left(\shat\right) \equiv \left|\mathcal{M}_{t}^{[k]}\right|^2 = \frac{\lambda_{q\phi}^2}{64\pi^2}\, \frac{M^4}{\shat^2}\, \left(\frac{\shat-4\BSMphimass^2}{\shat}\right)^{-1/2}\, \left[ \sum_{r=0}^k \frac{(-1)^{r+1}}{r+2} \left(\kappa_+^{r+2} - \kappa_-^{r+2}\right) \right]^2 \le 1 \,.
\label{eqn:hatOmegatEFT}
\end{equation}
We see that $\hat\Omega_{t,\,\text{EFT}}^{[k]}\!\left(\shat\right)$ goes to infinity as $\shat\to\infty$. Therefore, we can interpret the condition in \cref{eqn:hatOmegatEFT} as setting a perturbative unitarity cutoff for the parton-level center-of-mass energy $\sqrt{\shat}$. The precise value of this cutoff depends on the EFT truncation dimension $\Delta$, but it will be close to the new physics scale $M$.

\subsubsection*{$s\s$-Channel UV Theory}

Turning to the $s$-channel UV model defined in \cref{eqn:lagsUV}, the scattering amplitude is
\begin{equation}
\mathcal{A}_s = - \frac{\muq\muphi}{\shat - M^2 + iM\Gamma} -\lambda_q = -\lambda_q\, \frac{\shat + iM\Gamma}{\shat - M^2 + iM\Gamma} \,,
\label{eqn:scalarAs}
\end{equation}
where we have applied \cref{eq:lambdaqTune} to tune away the dimension-4 contribution.  Again following \cref{eqn:unitaritycondition}, the $s$-wave component is
\begin{equation}
\mathcal{M}_s = \left(\frac{\shat-4\BSMphimass^2}{\shat}\right)^{1/4} \frac{1}{8\pi}\, \mathcal{A}_s
= - \frac{\lambda_q}{8\pi}\s \frac{\shat + iM\Gamma}{\shat - M^2 + iM\Gamma}\! \left(\frac{\shat-4\BSMphimass^2}{\shat}\right)^{1/4} \,,
\label{eqn:Ms}
\end{equation}
which leads to the parton-level $s$-wave perturbative unitarity condition
\begin{equation}
\hat\Omega_{s,\,\text{UV}}\!\left(\shat\right) = \left|\mathcal{M}_s\right|^2 = \frac{\lambda_q^2}{64\pi^2}\, \frac{\shat^2 + M^2\Gamma^2}{\bigl(\shat-M^2\bigr)^2 + M^2\Gamma^2}\, \sqrt{\frac{\shat-4\BSMphimass^2}{\shat}} \le 1 \,.
\label{eqn:hatOmegasUV}
\end{equation}
The UV theory prediction is maximized on-resonance.  Hence, if the theory is free of perturbative unitarity violation when $\shat=M^2$, it will be free of perturbative unitarity violation for all $\shat$. To ensure this condition, we will set
\begin{align}
\lambda_q = 2 \,,
\label{eq:sChParam}
\end{align}
in the rest of this section (see \cref{fig:OmegasUV}).\footnote{We note that although $\hat\Omega_{s,\,\text{UV}}\!\left(\shat\right)$ hits 1 at $\shat=M^2$ for $\lambda_q=2$, imposing tree-level perturbative unitarity for $\sqrt{\shat}\gg M$ actually allows $\lambda_q$ to be as large as $8\pi$, similar to the $t$-channel case.}

\subsubsection*{$s\s$-Channel EFT}

We can obtain the corresponding EFT prediction $\hat\Omega_{s,\s\text{EFT}}(\shat)$ by repeating the above calculation for the EFT Lagrangian in \cref{eqn:lagsEFT}. Equivalently, we can just expand the UV result for the $s$-wave amplitude $\mathcal{M}_s$ given in \cref{eqn:Ms} in powers of $1/M^2$ (setting $\Gamma\to0$) up to some order $k$. This yields
\begin{equation}
\mathcal{M}^{[k]}_{s} = \frac{\lambda_q}{8\pi}\, \left(\frac{\shat-4\BSMphimass^2}{\shat}\right)^{1/4} \frac{\shat}{M^2}\, \sum_{r=0}^k \left(\frac{\shat}{M^2}\right)^r \,,
\end{equation}
which leads to the partonic $s$-channel EFT $s$-wave perturbative unitarity condition
\begin{equation}
\hat\Omega_{s,\s\text{EFT}}^{[k]}\!\left(\shat\right) \equiv \left|\mathcal{M}_{s}^{[k]}\right|^2 = \frac{\lambda_q^2}{64\pi^2}\, \sqrt{\frac{\shat-4\BSMphimass^2}{\shat}} \left[ \frac{\shat}{M^2}\, \sum_{r=0}^k \left(\frac{\shat}{M^2}\right)^r \right]^2 \le 1 \,.
\label{eqn:hatOmegasEFT}
\end{equation}
We see that $\hat\Omega_{s,\s\text{EFT}}^{[k]}\!\left(\shat\right)$ grows monotonically with $\shat$ (for $\shat\ge4\BSMphimass^2$) and goes to infinity as $\shat\to\infty$.  Therefore, the condition in~\cref{eqn:hatOmegasEFT}  places an upper bound on the parton-level center-of-mass energy $\sqrt{\shat}$, which is identified as the perturbative unitarity cutoff. Its precise value depends on the EFT truncation dimension $\Delta$, but it is always close to the new physics scale $M$.

\subsection{Hadronic Initial State}
\label{subsec:HadronUnitarity}

Next, we will derive the $s$-wave perturbative unitarity constraints for the hadronic scattering process in \cref{eqn:pairproduction}.  This requires generalizing the standard partial wave perturbative unitarity argument for pure initial quantum states to the case of mixed (or ensemble) initial quantum states.

To begin, recall that for a pure initial state $\ket{i}$, and a pure final state $\ket{f}\ne\ket{i}$, the usual perturbative unitarity condition reads
\begin{equation}
\hat\Omega_{i\to f} \equiv \left|\mathcal{M}_{i\to f}\right|^2 = \left| \mel**{f}{T}{i} \right|^2 \le 1 \,,
\end{equation}
where $T$ denotes the scattering operator $S=1+i\,T$. In order to generalize this to the case when the initial state is a mixed state, we rewrite the above condition using the density matrix of the (pure) initial state $\rho_i=\ketbra{i}{i}$:
\begin{equation}
\left| \mel**{f}{T}{i} \right|^2 \le 1
\qquad\Longleftrightarrow\qquad
\tr\big( \rho_i\, T^\dagger \ketbra{f}{f} T \big) \le 1 \,.
\label{eqn:purestatecondition}
\end{equation}
Next, we allow the initial state to be an ensemble, whose density matrix is given by
\begin{equation}
\rho_p = \sum_i\, p_i \ketbra{i}{i} = \sum_i\, p_i\, \rho_i \,,
\end{equation}
where $p_i\ge0$ are the coefficients (not necessarily normalized) of each pure-state density matrix $\ketbra{i}{i}$. In this case, if the condition in \cref{eqn:purestatecondition} holds for each pure-state, then it must be true that
\begin{equation}
\tr\left( \rho_p\, T^\dagger \ketbra{f}{f} T \right) = \sum_i\, p_i\s \tr\left( \rho_i\, T^\dagger \ketbra{f}{f} T \right) \le \sum_i\, p_i \,.
\label{eqn:mixedstateconditionRaw}
\end{equation}
One can further sharpen this condition by making use of the fact that certain selection rules can be imposed at the amplitude level. Suppose that for a specific final state $\ket{f}$, the amplitude can be nonzero only when the initial state $\ket{i}$ belongs to a subset $I$ of the ensemble
\begin{equation}
\left| \mel**{f}{T}{i} \right|^2 = \tr\big( \rho_i\, T^\dagger \ketbra{f}{f} T \big) = 0 \quad\text{for}\quad i \notin I \,.
\end{equation}
In this case, we can incorporate this effect into \cref{eqn:mixedstateconditionRaw}, which gives us
\begin{equation}
\tr\left( \rho_p\, T^\dagger \ketbra{f}{f} T \right) = \sum_{i\in I}\, p_i\s \tr\left( \rho_i\, T^\dagger \ketbra{f}{f} T \right) \le \sum_{i\in I}\, p_i \,.
\label{eqn:mixedstatecondition}
\end{equation}
For the application of interest here, we take the initial state to be the ensemble state formed by the pair of protons, and the final state to be $\ket{f}=\ket{\BSMphi\BSMphi^\dagger}$.  Then the left-hand side of \cref{eqn:mixedstatecondition} is nothing but the parton-level $\hat\Omega_{\phiq\phiq^\dagger\to\BSMphi\BSMphi^\dagger}\!\left(\shat\right)$ integrated over the parton distribution functions:
\begin{align}
\tr\left( \rho_p\, T^\dagger \ketbra{f}{f} T \right) &= \sum_i\, p_i\s \left| \mel**{f}{T}{i} \right|^2 \notag\\[5pt]
&= \sum_{\{q,\,\bar{q}\}\,\in\,p}\, \int_0^1 \dd x_1 \dd x_2\, \Big[ f_q(x_1) f_{\bar{q}}(x_2) + f_{\bar{q}}(x_1) f_q(x_2) \Big]\, \hat\Omega_{\phiq\phiq^\dagger\to\BSMphi\BSMphi^\dagger}\!\left(x_1 x_2 s\right) \,,
\label{eqn:LHS}
\end{align}
where the $f_q$ $(f_{\bar{q}})$ are the PDFs for quarks (anti-quarks), $x_1$ and $x_2$ are the corresponding momentum fractions, and we are suppressing the PDF dependence on the renormalization scale.

Since the parton-level $\hat\Omega_{\phiq\phiq^\dagger\to\BSMphi\BSMphi^\dagger}\!\left(\shat\right)$ only depends on the product
\begin{equation}
\tau \equiv x_1 x_2 = \shat/s \,,
\end{equation}
it is convenient to work with the parton luminosity function~\cite{Eichten:1984eu}
\begin{align}
L_{q\bar{q}}\left(\tau\right) &\equiv \int_0^1 \dd x_1 \dd x_2\, \Big[ f_q(x_1) f_{\bar{q}}(x_2) + f_{\bar{q}}(x_1) f_q(x_2) \Big]\, \delta\! \left( \tau - x_1 x_2 \right) \notag\\[5pt]
&= 2 \int_\tau^1 \dd x\, \frac{1}{x}\, f_q(x)\, f_{\bar{q}}(\tau/x) \,.
\label{eqn:PLFL}
\end{align}
This allows us to rewrite \cref{eqn:LHS} as
\begin{equation}
\tr\left( \rho_p\, T^\dagger \ketbra{f}{f} T \right) = \sum_{\{q,\,\bar{q}\}\,\in\,p}\, \int_{\tau_\phi}^1 \dd\tau\, L_{q\bar{q}}(\tau)\, \hat\Omega_{\phiq\phiq^\dagger\to\BSMphi\BSMphi^\dagger}\left(\shat = \tau s\right) \,,
\label{eq:tracepptopsipsi}
\end{equation}
where $\tau_\phi = 4\BSMphimass^2/s$ is the kinematic threshold for the pair production process at the parton level. On the other hand, the right-hand side of \cref{eqn:mixedstatecondition} can be written as
\begin{equation}
\sum_i\, p_i = \sum_{\{q,\,\bar{q}\}\,\in\,p}\, \int_{\tau_\phi}^1 \dd\tau\, L_{q\bar{q}}(\tau) \,.
\end{equation}
(Note that we are suppressing the explicit dependence on the selection rule in the sum, see \cref{eqn:mixedstatecondition}.) Therefore, we obtain the $s$-wave perturbative unitarity condition for the hadronic scattering process $p\s p\to\BSMphi\s\BSMphi^\dagger$:
\begin{equation}
\Omega_{pp\to\BSMphi\BSMphi^\dagger}\left(s\right) \equiv \frac{\sum_{\{q,\,\bar{q}\}\,\in\,p}\, \int_{\tau_\phi}^1 \dd\tau\, L_{q\bar{q}}(\tau)\, \hat\Omega_{\phiq\phiq^\dagger\to\BSMphi\BSMphi^\dagger}\!\left(\shat = \tau s\right)}{\sum_{\{q,\,\bar{q}\}\,\in\,p}\, \int_{\tau_\phi}^1 \dd\tau\, L_{q\bar{q}}(\tau)}  \le 1 \,.
\label{eqn:HadronicUnitarityBound}
\end{equation}
This result applies to the case of either scalar or fermionic initial state partons.

\subsection{Unitarity and EFT Truncation}
\label{subsec:UnitarityResults}

Now that we have a hadronic formalism for the $s$-wave perturbative unitarity condition $\Omega \le 1$, we can investigate its implications when interpreted as an EFT validity test. We will show results for both the $t$-channel and $s$-channel models, finding that they yield similar results. Note that to better understand the perturbative unitarity constraints, we will be varying the partonic (hadronic) center-of-mass energy $\sqrt{\shat}$ ($\sqrt{s}$) in what follows. In particular, $\sqrt{s}$ will not be fixed to 14 TeV. We therefore introduce a notation $E_\text{cm}$ that denotes this varying center-of-mass energy to avoid confusion.

For the numerical evaluation, we use the CT10 PDFs~\cite{Lai:2010vv} and set the renormalization scale to 10 TeV for convenience.\footnote{The PDFs minimally change as we vary the renormalization scale from 3 to 100 TeV.}$^{,}$\footnote{For certain values of $M$ and $\Delta$, efficiently performing the numerical integral over the PDFs is nontrivial. To overcome this challenge, we implemented the adaptive Simpson's method along with some carefully chosen variable changes.} We choose $m_{\BSMphi}=1$ TeV as a benchmark value, but we emphasize that choosing other values of $\BSMphimass$ will yield results with the same qualitative features. To support this claim, we have included a set of results for the case $\BSMphimass=10$ GeV in \cref{appsec:BSMpsimassLow}.

\subsubsection*{$t\s$-Channel UV Theory}

\begin{figure}[h!]
\centering
\includegraphics[width=0.4\textwidth]{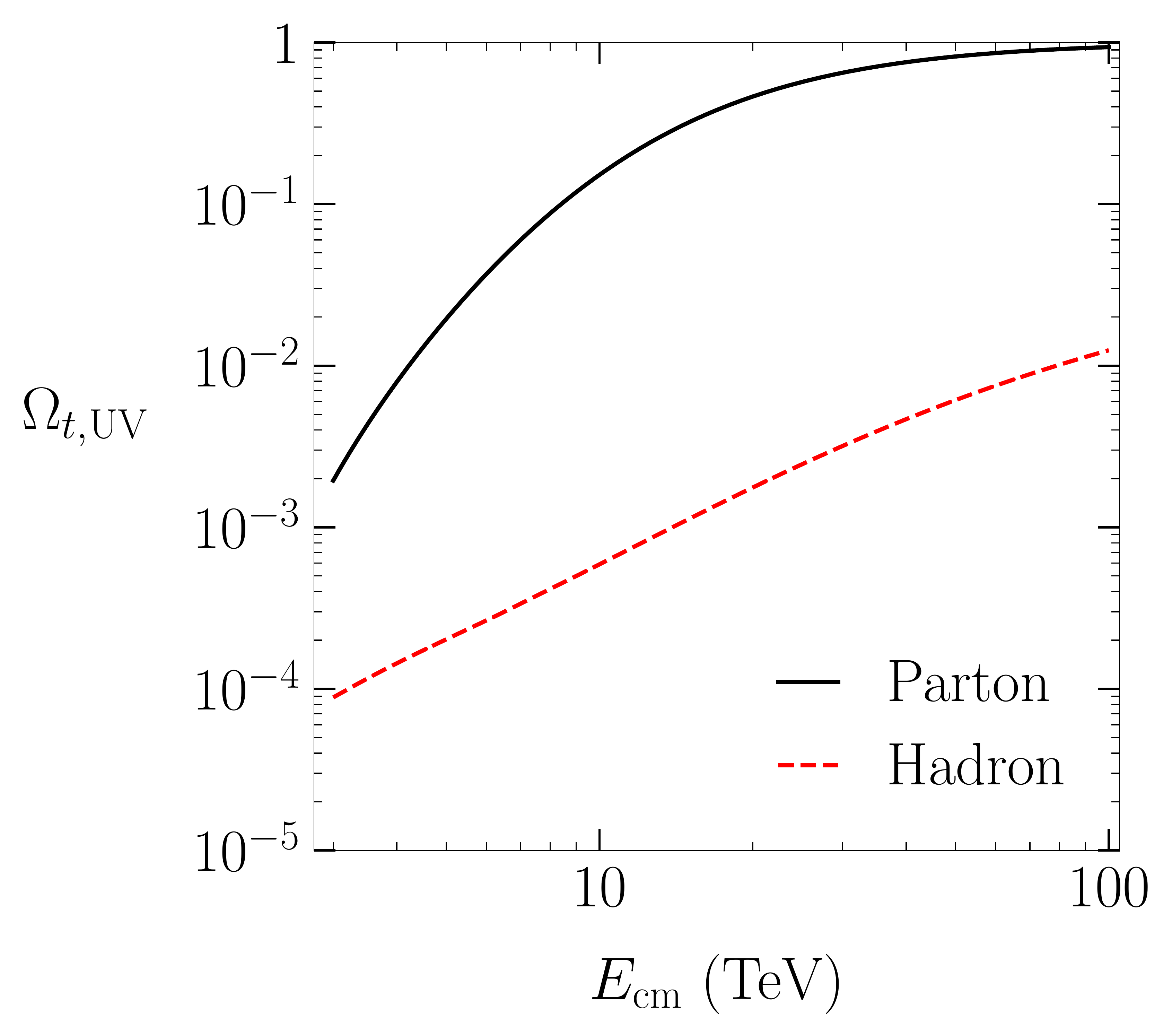}
\caption{$\Omega_{t,\s\text{UV}}$ computed using the $t$-channel UV model as a function of the center-of-mass energy $E_\text{cm}$ for parton and hadron initial states. This shows that the UV theory is free of perturbative unitarity violation, when the couplings are taken to be $\lambda_{q\phi} = \left|\muqphi\right|^2/M^2 = 8\pi$.}\label{fig:OmegatUV}
\end{figure}

We begin by investigating the validity of the UV theory by checking the $s$-wave perturbative unitarity condition for both the partonic and hadronic cases. In \cref{fig:OmegatUV}, we plot typical curves of $\Omega_{t,\s\text{UV}}$ as a function of the center-of-mass energy $E_\text{cm}$.  We see that with our choice of the coupling $\lambda_{q\phi} = \left|\muqphi\right|^2/M^2 = 8\pi$ (see~\cref{eq:lambdaQPhi,eq:tChParam}), the UV theory is free of perturbative unitarity violation. Specifically, $\Omega_{t,\s\text{UV}} < 1$ across the $E_\text{cm}$ range of interest, in both the partonic initial state and the hadronic initial state cases.

\subsubsection*{$t\s$-Channel EFT}

\begin{figure}[h!]
\centering
\includegraphics[width=0.9\textwidth]{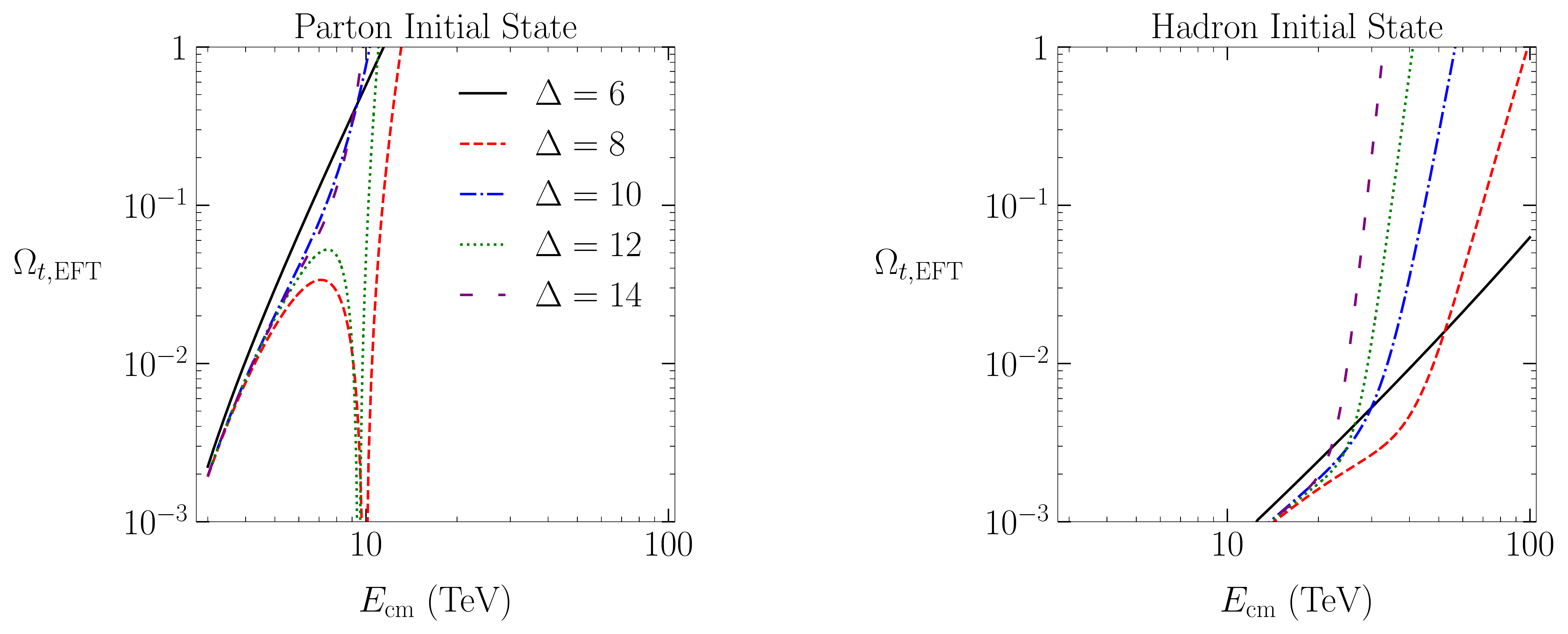}
\caption{$\Omega_{t,\,\text{EFT}}$ computed using the EFT expansion of the $t$-channel model as a function of the center-of-mass energy $E_\text{cm}$, for low choices of the truncation dimension $\Delta=6+2k$. For the Partonic Initial State case [left], when $\Delta>0$, $\Omega_{t,\,\text{EFT}}$ grows at large $E_\text{cm}$ and approaches infinity as $E_\text{cm}\to\infty$. This tells us there will be a perturbative unitarity cutoff for a critical value of $E_\text{cm}$. In the Hadronic Initial State case [right], the growth of $\Omega_{t,\,\text{EFT}}$ is significantly delayed as compared to the partonic case.}\label{fig:OmegatEFTs}
\end{figure}

\begin{figure}[h!]
\centering
\vspace{60pt}
\includegraphics[width=0.9\textwidth]{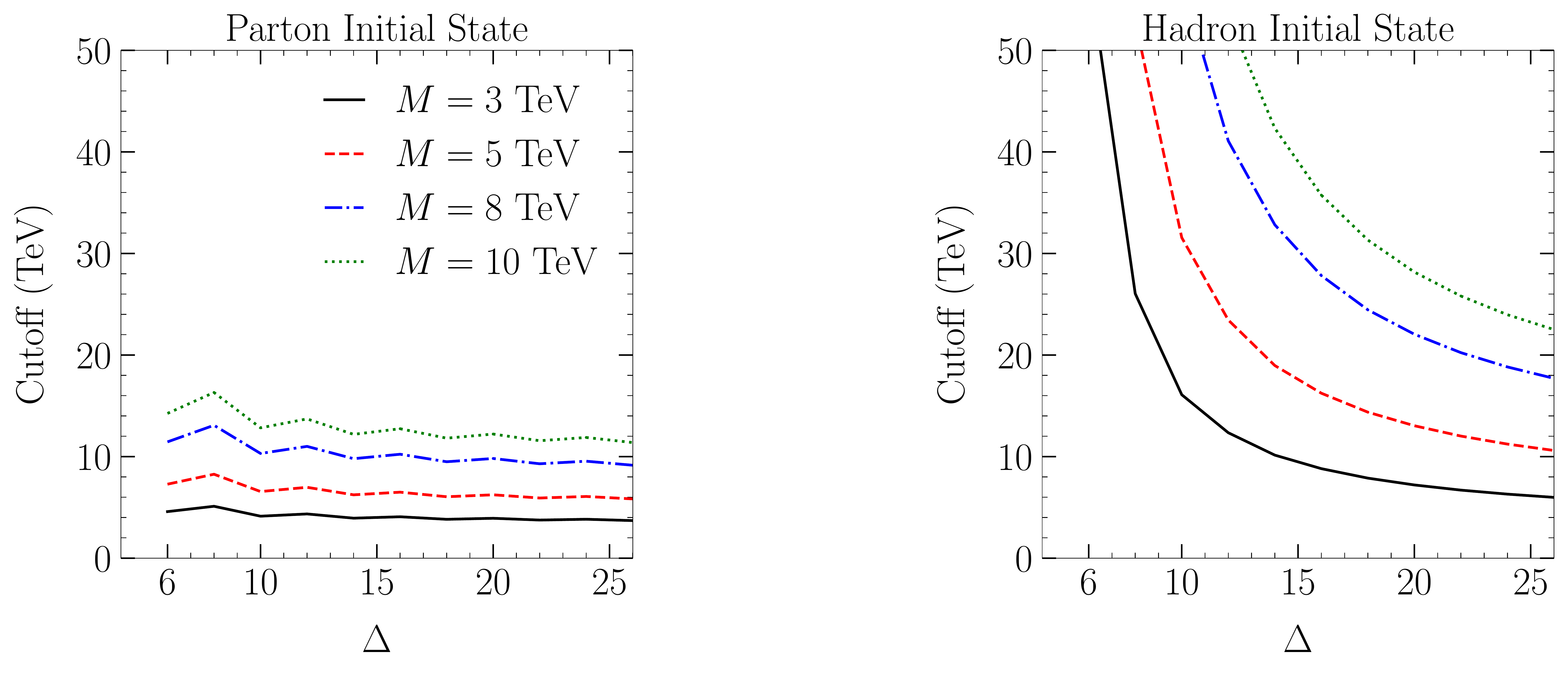}
\caption{The perturbative unitarity cutoff on $E_\text{cm}$ as a function of the EFT truncation dimension $\Delta$ for the $t$-channel model, derived using \cref{UnitarityCondition}. In the Partonic Initial State case [left], the perturbative unitarity cutoffs are more severe than for the Hadronic Initial State case [right], although the PDF effects do not fully remove the bounds.
}\label{fig:OmegatEcmBound}
\end{figure}

Next, we investigate the consequences for the EFTs. In~\cref{fig:OmegatEFTs}, we provide typical curves of  $\Omega_{t,\,\text{EFT}}$ as a function of the center-of-mass energy $E_\text{cm}$. We see that $\Omega_{t,\,\text{EFT}}$ becomes larger than 1 in the $E_\text{cm}$ range of interest, indicating a perturbative unitarity cutoff on $E_\text{cm}$.\footnote{The growth with $E_\text{cm}$ is not monotonic due to the $(-1)^r$ factor in \cref{eqn:hatOmegatEFT}, which comes from the fact that $\that<0$.} Comparing the partonic initial state case [left] to the hadronic initial state case [right], we see that the growth of $\Omega_{t,\,\text{EFT}}$ is significantly delayed by the PDF suppression of high-energy partons. Note that the curves are not flattened by the PDFs.  This implies that although the perturbative unitarity cutoff on $E_\text{cm}$ will be pushed significantly higher in the hadronic case, it will not be eliminated, as explicitly verified by \cref{fig:OmegatEcmBound}. Moreover, the cutoff on $E_\text{cm}$ is reduced as we increase the EFT truncation dimension $\Delta$, and eventually approaches $M$ in the large $\Delta$ limit. This implies that for $E_\text{cm}>M$, perturbative unitarity violation is guaranteed if one keeps including more operators beyond a critical truncation dimension.

We conjecture that this is a generic feature of EFTs used in collider searches.  This motivates adopting the following criterion for when an EFT is invalid:
\begin{align}
\text{The EFT truncated to dimension $\Delta = 6+2k$ is invalid if}\;\; \Omega_{\text{EFT}}^{[k]}(s) > 1 \,.
\label{UnitarityCondition}
\end{align}
Note that this does not guarantee that the EFT is a good description of some underlying UV physics outside of the region deemed invalid by this criterion.

\begin{figure}[t!]
\centering
\includegraphics[width=0.4\textwidth]{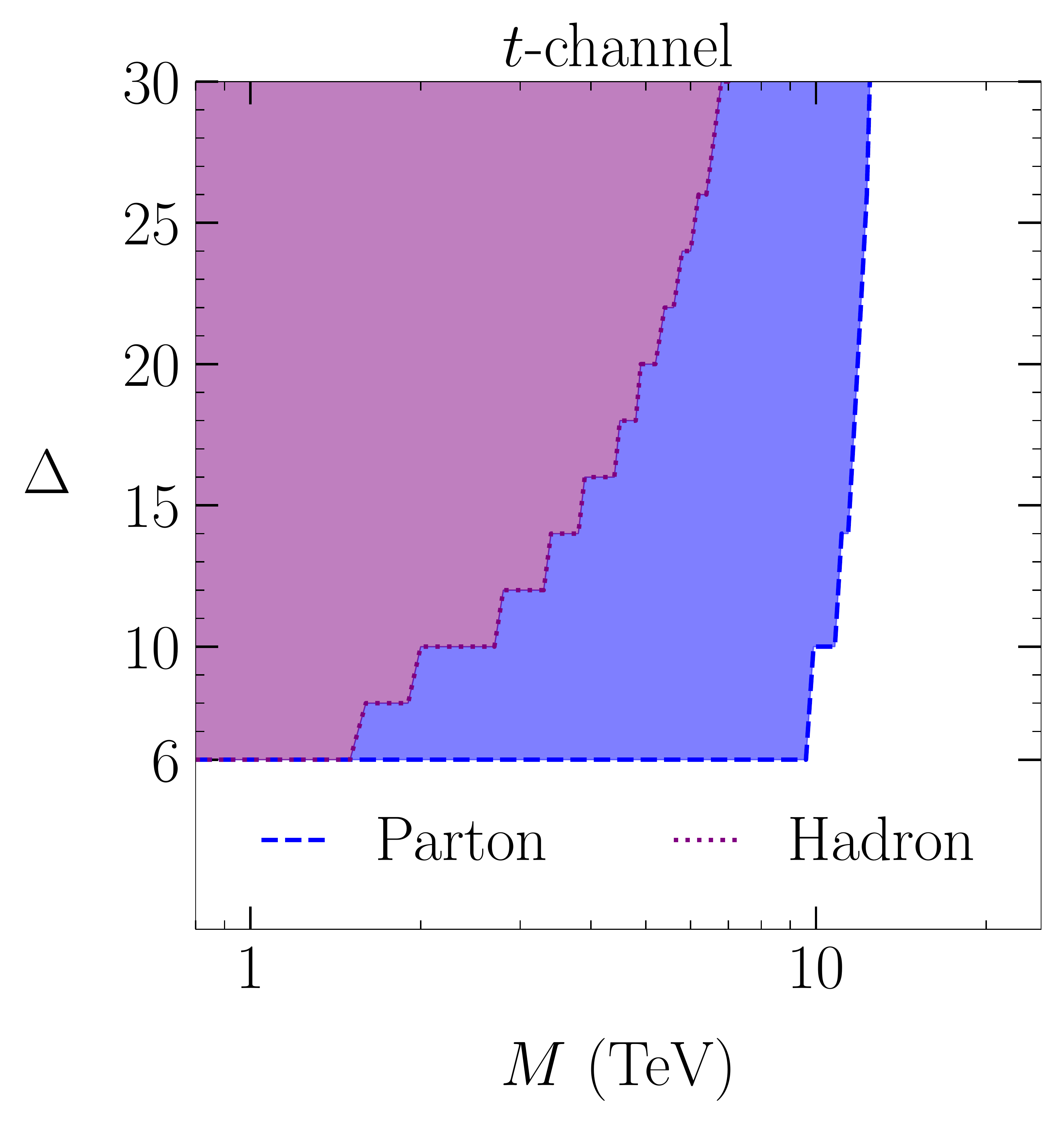}
\caption{The shaded region shows the parameter space where the EFT is invalid for the $t$-channel model in the plane of the EFT truncation dimension $\Delta$ versus the BSM scale $M$ with $\lambda_{q\phi}=8\pi$. In the Partonic (Hadronic) Initial State case, we take $\sqrt{\shat}$ ($\sqrt{s}$) $=14$ TeV. The inclusion of PDF effects opens up a region of potentially viable parameter space.} \label{fig:OmegatDimBound}
\end{figure}

In \cref{fig:OmegatDimBound}, we plot the invalid region in the $(M, \Delta)$ parameter space obtained by applying this criterion to the $t$-channel model. We see that going from partons to hadrons opens up significant parameter space for which the EFT could be a valid description.  This tells us that perturbative unitarity arguments for the invalidity of EFT analyses performed at hadron colliders in the parameter space where $M < \sqrt{s}$ should incorporate PDF effects when the search region is sufficiently inclusive.

\subsubsection*{$s\s$-Channel UV Theory}

\begin{figure}[h!]
\centering
\includegraphics[width=0.4\textwidth]{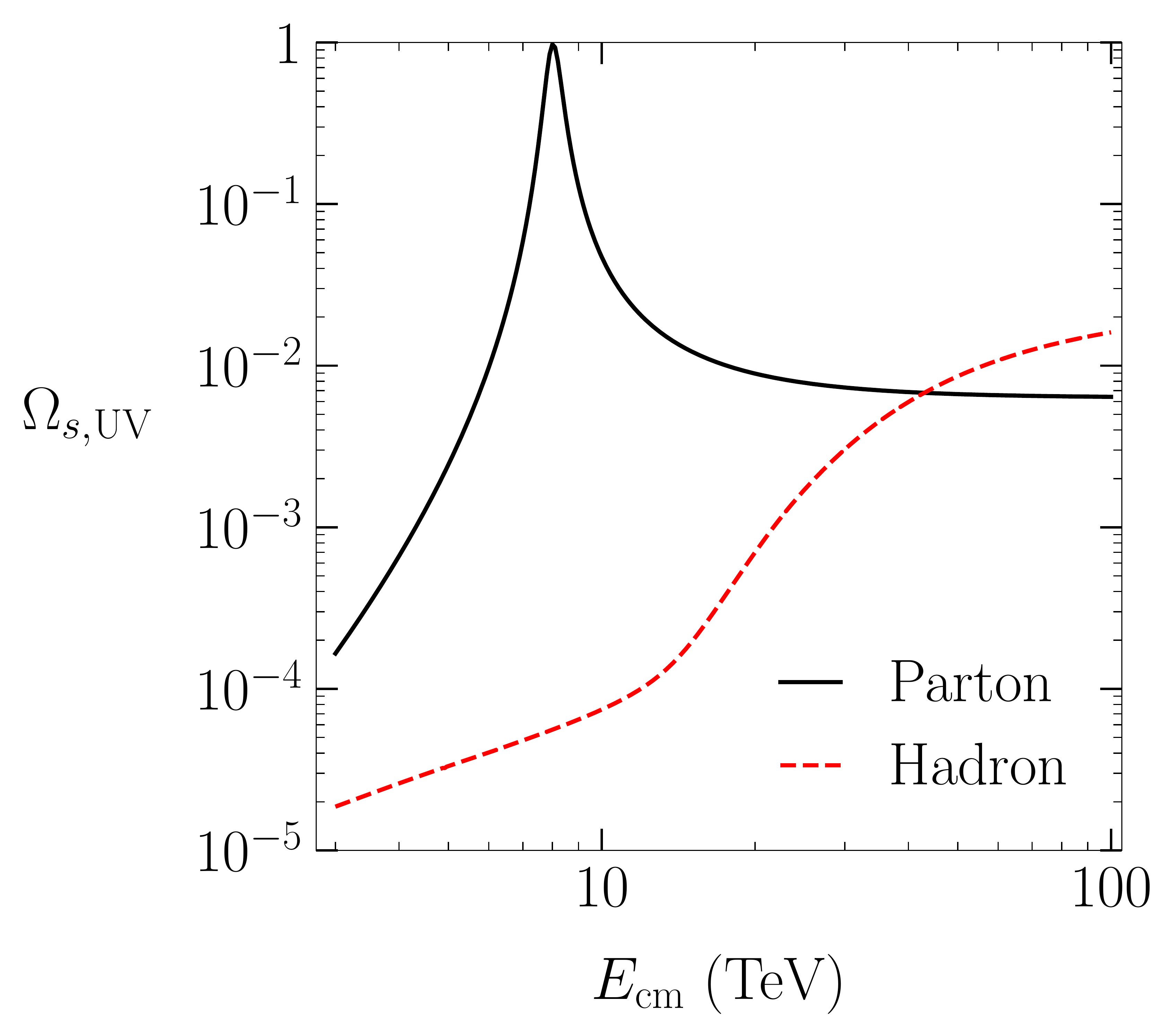}
\caption{$\Omega_{s,\,\text{UV}}$ computed using the $s$-channel UV model as a function of the center-of-mass energy $E_\text{cm}$. This shows that the UV theory is free of perturbative unitarity violation, when the couplings are taken to be $\lambda_q=\muq\muphi/M^2=2$.}
\label{fig:OmegasUV}
\end{figure}

The $s$-channel model yields qualitatively similar results to those we found for the $t$-channel case. As before, we begin by checking the $s$-wave perturbative unitarity of the UV theory. In \cref{fig:OmegasUV}, we plot $\Omega_{s,\,\text{UV}}$ as a function of the center-of-mass energy $E_\text{cm}$. We see that for our choice of the couplings $\lambda_q=\muq\muphi/M^2=2$ (see \cref{eq:lambdaqTune,eq:sChParam}), the UV theory is free of perturbative unitarity violation; $\Omega_{s,\,\text{UV}} < 1$ across the $E_\text{cm}$ range of interest at both the parton and hadron level.  The resonance feature is clear when varying $E_\text{cm}$ for the partonic case, but it is smeared out by PDF effects for the hadronic case.

\subsubsection*{$s\s$-Channel EFT}

\begin{figure}[h!]
\centering
\includegraphics[width=0.4\textwidth]{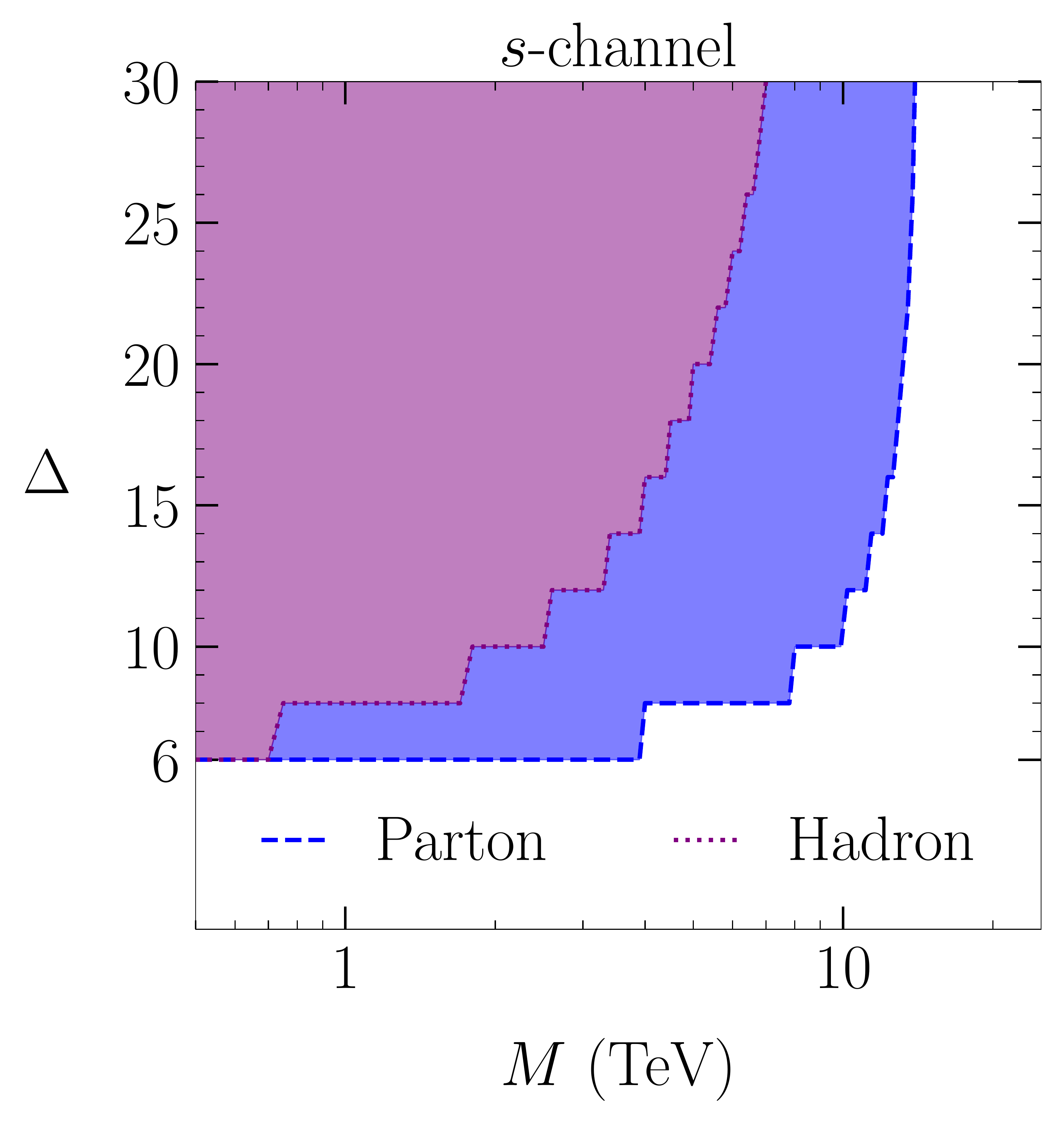}
\caption{The shaded region shows the parameter space in the plane of the EFT truncation dimension $\Delta$ versus the BSM scale $M$ with $\lambda_q=2$, where the EFT is deemed invalid using the criterion in \cref{UnitarityCondition} for the $s$-channel model. In the Partonic (Hadronic) Initial State case, we take $\sqrt{\shat}$ ($\sqrt{s}$) $=14$ TeV. The inclusion of PDF effects opens up a region of potentially viable parameter space.} \label{fig:OmegasDimBound}
\end{figure}

Switching to the EFTs, the requirement of $s$-wave perturbative unitarity will again tell us that the EFT becomes invalid for some large $E_\text{cm}$.  As with the $t$-channel scenario, the growth of $\Omega_{s,\s\text{EFT}}$ is significantly delayed by the PDF suppression of high-energy partons in the hadronic initial state case.  In \cref{fig:OmegasDimBound}, we plot the invalid region in the $(M, \Delta)$ parameter space obtained by applying \cref{UnitarityCondition} to the $s$-channel model. We again see that going from partons to hadrons opens up significant parameter space for which the EFT could be a valid description.  Note we have taken $\Gamma \to 0$ in the EFT expansion when deriving these bounds.

\clearpage

\section{Interpreting Unitarity Violation}
\label{sec:Xsections}

So far, we have simply explored the impact of PDFs on partial wave perturbative unitarity bounds.  In particular, we showed the quantitative impact that PDF suppression has on the high energy growth of EFT amplitudes for sufficiently inclusive search regions.  This suppression postpones the scale of perturbative unitarity violation, thereby potentially opening up parameter space with $M < \sqrt{s}$ where the EFT could be a useful description.  The goal of this section is to interpret these results by comparing them against the predictions for a physical observable.

We will continue to focus on the simple 2-to-2 scattering process in \cref{eqn:pairproduction}. We compare the predictions for its cross section $\sigma$ as derived from the UV theory and the EFT as we vary the truncation dimension $\Delta$ and the mediator mass $M$ against the invalid regions derived in the previous section.  For our purposes here, a ``valid'' EFT is one that
\begin{enumerate}[label=(\roman*)]
\setlength\itemsep{1pt}
\item reproduces the full theory cross section to a reasonable approximation, and
\item converges toward the full theory result as $\Delta$ is increased.
\end{enumerate}
We will show that valid EFTs exist in the region opened up by PDF effects.

To determine the hadronic pair production cross section $\sigma_{pp\to\BSMphi\BSMphi^\dagger}(s)$ from the corresponding partonic one $\hat\sigma_{\phiq\phiq^\dagger\to\BSMphi\BSMphi^\dagger}(\shat)$, we integrate the partonic cross section over the parton distribution functions using the standard formula
\begin{equation}
\sigma_{pp\to\BSMphi\BSMphi^\dag}(s) = \sum_{\{q,\,\bar{q}\}\,\in\,p}\, \int_{\tau_\phi}^1 \dd\tau\, L_{q\bar{q}}(\tau)\, \hat\sigma_{\phiq\phiq^\dagger\to\BSMphi\BSMphi^\dagger}\!\left(\shat = \tau s\right) \,,
\label{eq:sigmapptopsipsi}
\end{equation}
where we have used the parton luminosity function defined in \cref{eqn:PLFL}, and the lower bound on the integral $\tau_\phi = 4\BSMphimass^2/s$ is the kinematic threshold for the pair production process at the parton level.

We are interested in varying the new physics scale $M$ while investigating to what extent
\begin{equation}
\sigma_\text{EFT}^{[k]} \equiv \sum_{r=0}^k \sigma_{\text{EFT}}^{(r)} \stackrel{?}{\simeq} \sigma_\text{UV} \,,
\label{eqn:Question}
\end{equation}
where we are defining the notation $\sigma^{[k]}$ to distinguish the cross section that includes the sum of EFT contributions up to order $k$ ($\Delta=6+2k$; see~\cref{eq:Delta6}), from the contribution of an individual term $\sigma^{(r)}$.  To this end, \cref{subsec:PartonXsections} provides the predictions for the parton-level cross sections $\hat\sigma_{\phiq\phiq^\dag\to\BSMphi\BSMphi^\dag}(\shat)$ for the UV theories and the EFTs detailed in \cref{sec:Models}. The main results of this section are given in \cref{subsec:XsectionResults}, where we investigate the question posed in \cref{eqn:Question} by comparing the numerical results for $\sigma_\text{EFT}$ and $\sigma_\text{UV}$ for different choices of $M$ and $\Delta$, and use these results to interpret the perturbative partial-wave unitarity results of the previous section.

In the parameter space $M < \sqrt{s}$, we will show that in the limit $\Delta\to\infty$, the EFT expansion of the cross section is not a convergent series.  This implies that one cannot blindly increase the truncation dimension $\Delta$ to achieve an arbitrarily good approximation of the underlying UV physics. Nevertheless, thanks to the PDF suppression of high-energy partons, when $\Delta$ is small, the relative error between the UV and EFT predictions actually decreases with $\Delta$, as if it were a convergent series. Then for $\Delta$ larger than a critical value $\D$, the error will begin to grow with $\Delta$.  This tells us that an EFT analysis performed at low orders can provide an adequate approximation of the underlying UV physics that improves with $\Delta$, even when $M < \sqrt{s}$.

\subsection{Partonic Initial State Cross Sections}
\label{subsec:PartonXsections}

The parton-level cross sections $\hat\sigma_{\phiq\phiq^\dagger\to\BSMphi\BSMphi^\dagger}(\shat)$ can be computed from the amplitudes derived in the previous section.

\subsubsection*{$t$-channel UV Theory}

We begin with the $t$-channel UV model defined in \cref{eqn:lagtUV}. The 2-to-2 scattering amplitude is given in \cref{eqn:scalarAt}.
This yields the color averaged squared amplitude
\begin{equation}
\overline{|\mathcal{A}_t|^2} = \frac{1}{3}\s \lambda_{q\phi}^2\s \frac{\that^2}{\bigl(\that-M^2\bigr)^2} \,.
\label{eqn:Ataverage}
\end{equation}
Using the kinematic relation in \cref{eqn:thatintheta}, we can integrate over the scattering angle to derive the parton-level total cross section
\begin{align}
\hat\sigma_{t,\s\text{UV}}(\shat) &= 2\pi \int_{-1}^1 \dd (\cos\theta) \frac{1}{64\pi^2 \shat}\s\overline{|\mathcal{A}_t|^2}\, \sqrt{\frac{\shat-4\BSMphimass^2}{\shat}} \notag\\[8pt]
&= \frac{\lambda_{q\phi}^2}{48\pi}\s \frac{M^2}{\shat^2}\! \left[ \kappa_+ - \kappa_- + \frac{\kappa_+ - \kappa_-}{(1+\kappa_+)(1+\kappa_-)} -2 \log\frac{1+\kappa_+}{1+\kappa_-} \right] \,,
\label{eqn:hatsigmatUV}
\end{align}
where $\kappa_\pm$ is defined in \cref{eqn:auxiliary}, and we are assuming that the initial state scalar quarks are massless.

\subsubsection*{$t$-channel EFT}

The EFT predictions for the $t$-channel production cross section can be obtained by repeating the above calculation with the Lagrangian defined in \cref{eqn:lagtEFT}. This amounts to expanding the UV result in \cref{eqn:hatsigmatUV} in powers of $1/M^2$ (encoded by the $\kappa_\pm$ dependence, see \cref{eqn:auxiliary}) and truncating the expansion at some EFT order $k$:
\begin{subequations}\label{eqn:hatsigmatEFT}
\begin{align}
\hat\sigma_{t,\,\text{EFT}}^{[k]}(\shat) &= \sum_{r=0}^k \hat\sigma_{t,\,\text{EFT}}^{(r)}(\shat) \,,\\[8pt]
\hat\sigma_{t,\,\text{EFT}}^{(r)}(\shat) &= \frac{\lambda_{q\phi}^2}{48\pi}\s \frac{M^2}{\shat^2}\s \frac{r+1}{r+3}\s (-1)^r\! \left( \kappa_+^{r+3} - \kappa_-^{r+3} \right) \,.
\label{eqn:hatsigmatEFTr}
\end{align}
\end{subequations}

\subsubsection*{$s$-channel UV Theory}

Now we turn to the $s$-channel UV model defined in \cref{eqn:lagsUV}. The 2-to-2 scattering amplitude is given in \cref{eqn:scalarAs}.
The color averaged squared amplitude is then
\begin{equation}
\overline{|\mathcal{A}_s|^2} = \frac13\s \lambda_q^2\s \frac{\shat^2 + M^2\Gamma^2}{\bigl(\shat-M^2\bigr)^2 + M^2\Gamma^2} \,,
\label{eqn:Asaverage}
\end{equation}
which leads to the parton-level cross section
\begin{equation}
\hat\sigma_{s,\,\text{UV}}(\shat) = \frac{1}{16\pi\s \shat}\, \overline{|\mathcal{A}_s|^2}\, \sqrt{\frac{\shat-4\BSMphimass^2}{\shat}}
= \frac{\lambda_q^2}{48\pi}\s \frac{1}{\shat}\s \frac{\shat^2 + M^2\Gamma^2}{\bigl(\shat-M^2\bigr)^2 + M^2\Gamma^2}\s\sqrt{\frac{\shat-4\BSMphimass^2}{\shat}} \,,
\label{eqn:hatsigmasUV}
\end{equation}
where we are treating the initial state quarks as massless.  For the numerics that follow, we will always take $\Gamma=M/(4\pi)$ for simplicity.

\subsubsection*{$s$-channel EFT}

To work out the EFT predictions for the $s$-channel production cross section, we can repeat the above calculation with the Lagrangian given in \cref{eqn:lagsEFT}, with the width effects incorporated. Equivalently, one can expand the UV result in \cref{eqn:hatsigmasUV} in powers of $1/M^2$ and truncating the expansion at some order $k$. This yields
\begin{subequations}\label{eqn:hatsigmasEFT}
\begin{align}
\hat\sigma_{s,\s\text{EFT}}^{[k]}(\shat) &= \sum_{r=-2}^k \hat\sigma_{s,\s\text{EFT}}^{(r)}(\shat) \,, \label{eqn:hatsigmasEFTk} \\[5pt]
\hat\sigma_{s,\s\text{EFT}}^{(r)}(\shat) &= \frac{\lambda_q^2}{48\pi M^2}\s \sqrt{\frac{\shat-4\BSMphimass^2}{\shat}}\s c_r(\Gamma/M) \left(\frac{\shat}{M^2}\right)^{r+1} \,,
\label{eqn:hatsigmasEFTr}
\end{align}
\end{subequations}
\clearpage
\noindent where the coefficient is defined as
\begin{equation}
c_{r-2}(\Gamma/M) \equiv \frac{1}{r!}\! \left(\frac{\partial}{\partial x}\right)^r\s\! \left[\frac{x^2 + (\Gamma/M)^2}{(1-x)^2 + (\Gamma/M)^2}\right] \Biggr|_{x=0} \,.
\label{eqn:cr}
\end{equation}
Note that the sum in \cref{eqn:hatsigmasEFTk} starts with $r=-2$ in order to capture the width effects $\Gamma/M\ne0$ (which technically only appear at loop level). In the zero width limit, the $r=-2$ and $r=-1$ terms would vanish ($c_{-2}=c_{-1}=0$), because the expression in the square bracket in \cref{eqn:cr} would have a Taylor expansion that starts with $x^2$.  The reason we are incorporating the width effects in the EFT matching is that they will be important for properly examining the question posed in \cref{eqn:Question} when one goes to sufficiently high truncation dimension.  We will explore the impact of this ``width improved matching'' when we compare \cref{fig:sigmasErrorCollect,fig:sigmasErrorCollect_NoWidth} below.

\subsection{Evidence for EFT Validity}
\label{subsec:XsectionResults}

With the cross sections $\sigma_\text{UV}$ and $\sigma_\text{EFT}$ in hand, we can now turn to answering the question raised in \cref{eqn:Question}. Specifically, we investigate the behavior of the relative error as a function of the truncation dimension $\Delta = 6 +2k$:
\begin{equation}
\text{Relative Error} \equiv \frac{\sigma_{\text{EFT}}^{[k]}}{\sigma_{\text{UV}}} - 1 \,,
\label{eqn:Errors}
\end{equation}
which provides a proxy for the question of EFT validity. Another useful quantity for exploring this question is the ``power counting uncertainty'' on the EFT prediction, which we will compute using
\begin{equation}
\text{Power Counting Uncertainty} \equiv \left| \frac{\sigma_\text{EFT}^{(k+1)}}{\sigma_\text{EFT}^{[k]}}\right| \,.
\label{eqn:Uncertainty}
\end{equation}
This captures the fact that the EFT is an approximation of the full theory, and this power counting uncertainty provides an indication for the level of confidence one should have when using the EFT prediction.

We will provide results for the $t$-channel and $s$-channel models separately; while they are qualitatively similar to each other, we will highlight some interesting differences in the details. For the numerical results that follow, we again use the CT10 PDFs~\cite{Lai:2010vv} and set the renormalization scale to 10 TeV for convenience. For the BSM singlet mass, we stick to our benchmark value $\BSMphimass=1$ TeV; other values would yield results with the same qualitative feature, as supported by \cref{appsec:BSMpsimassLow}, where we show some results with $m_{\BSMphi}=10$ GeV.

\subsubsection{$t$-channel Results}

\begin{figure}[t]
\centering
\includegraphics[width=0.9\textwidth]{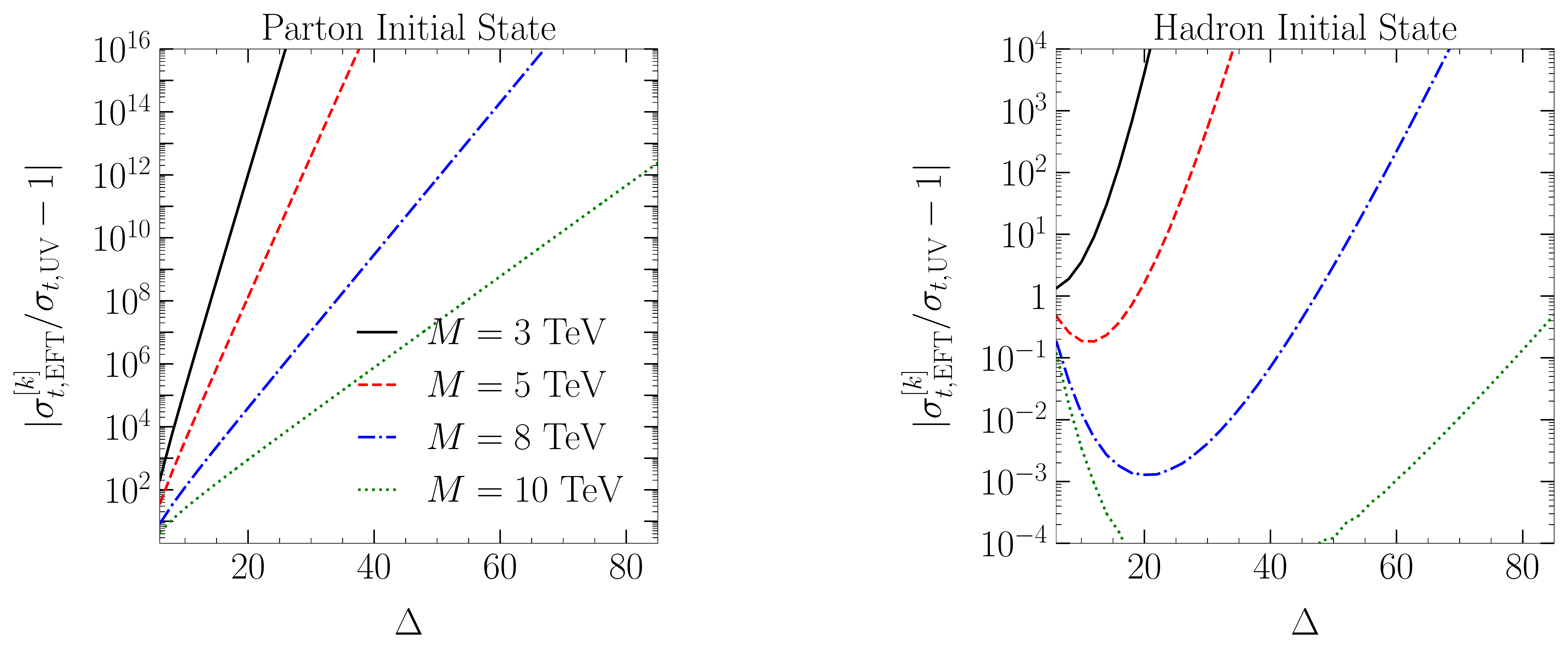}
\caption{The absolute value of the relative error (see \cref{eqn:Errors}) computed for the $t$-channel model as a function of the EFT truncation dimension $\Delta$. For the ``Partonic Initial State'' case [left], we present curves for $M<\sqrt{\shat}=14$ TeV, which show that the error grows monotonically as $\Delta$ is increased.  In the ``Hadronic Initial State'' case [right], we present curves for $M<\sqrt{s}=14$ TeV, which show that the EFT approximation improves for small values of $\Delta$, but then the error begins to grow for $\Delta > \D$.}\label{fig:sigmatAbsErrorCollect}
\end{figure}

\begin{figure}[t]
\centering\vspace{0.8cm}
\includegraphics[width=1.0\textwidth]{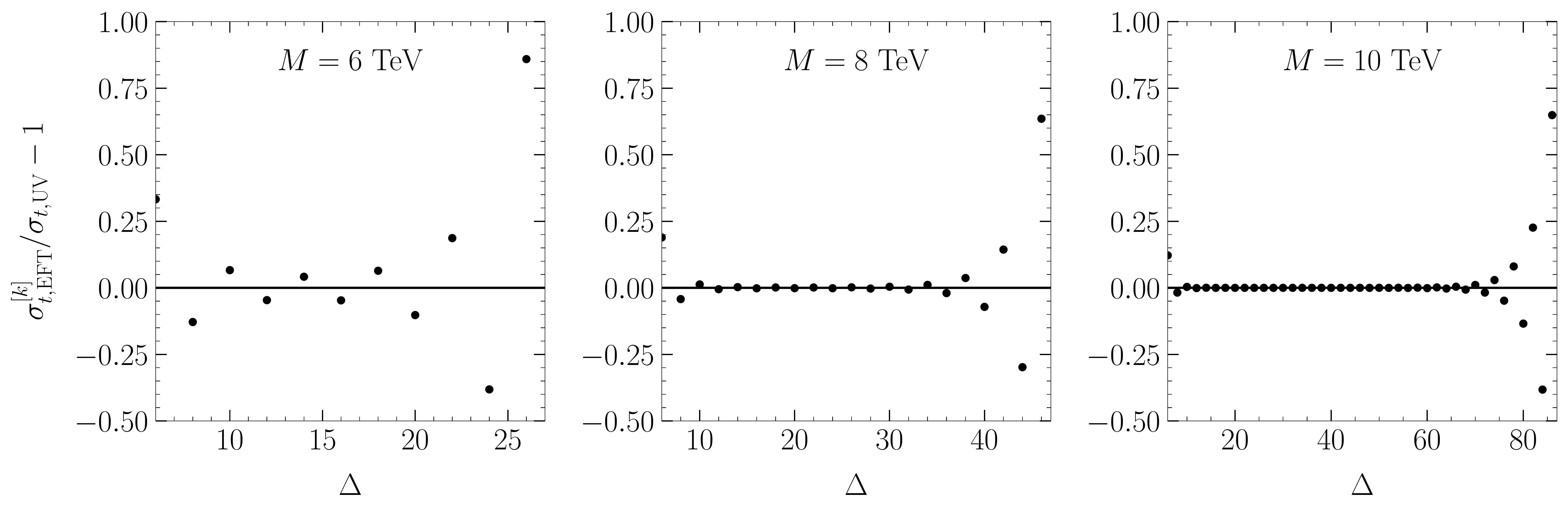}
\caption{The relative error (see \cref{eqn:Errors}) as computed for the $t$-channel model in the ``Hadronic Initial State'' case for $M<\sqrt{s}=14$ TeV.}\label{fig:sigmatErrorCollect}
\end{figure}

We plot the absolute value of the relative error in \cref{fig:sigmatAbsErrorCollect}. In the parton case with $M<\sqrt{\shat}=14$ TeV, the error grows monotonically with $\Delta$, meaning that the EFT approximation keeps getting worse as $\Delta$ is increased. This is exactly the expected behavior, since this is effectively attempting to do an expansion when the relevant parameter $\hat{s}/M^2 > 1$, see \cref{eqn:hatsigmatEFTr}. For contrast, in the hadron case (now with $M<\sqrt{s}=14$ TeV), we find that the error \emph{decreases} with $\Delta$ for small values of $\Delta$, but turns around at some point and starts increasing at larger $\Delta$. We summarize this intriguing behavior of the EFT results:
\begin{itemize}
\setlength\itemsep{1pt}
\item The hadronic EFT expansion appears to be converging at lower orders:~we see the EFT approximation improving before hitting a critical value $\D$.
\item The hadronic EFT expansion series does not converge absolutely:~it becomes arbitrarily poor at sufficiently large $\Delta$.
\end{itemize}
In order to illuminate this appearing-to-be converging feature, we provide \cref{fig:sigmatErrorCollect}, which shows typical curves of the relative error without taking the absolute value to highlight how $\D$ is approached.\footnote{Note that the $t$-channel result alternates in sign, due to the $(-1)^r$ factor in \cref{eqn:hatsigmatEFTr}, which appears since $\that<0$.}$^{,}$\footnote{This behavior of the EFT validity is very similar, in appearance, to the validity of the perturbation expansion series for the scattering matrix at low orders.}

\begin{figure}[t]
\centering
\includegraphics[width=0.9\textwidth]{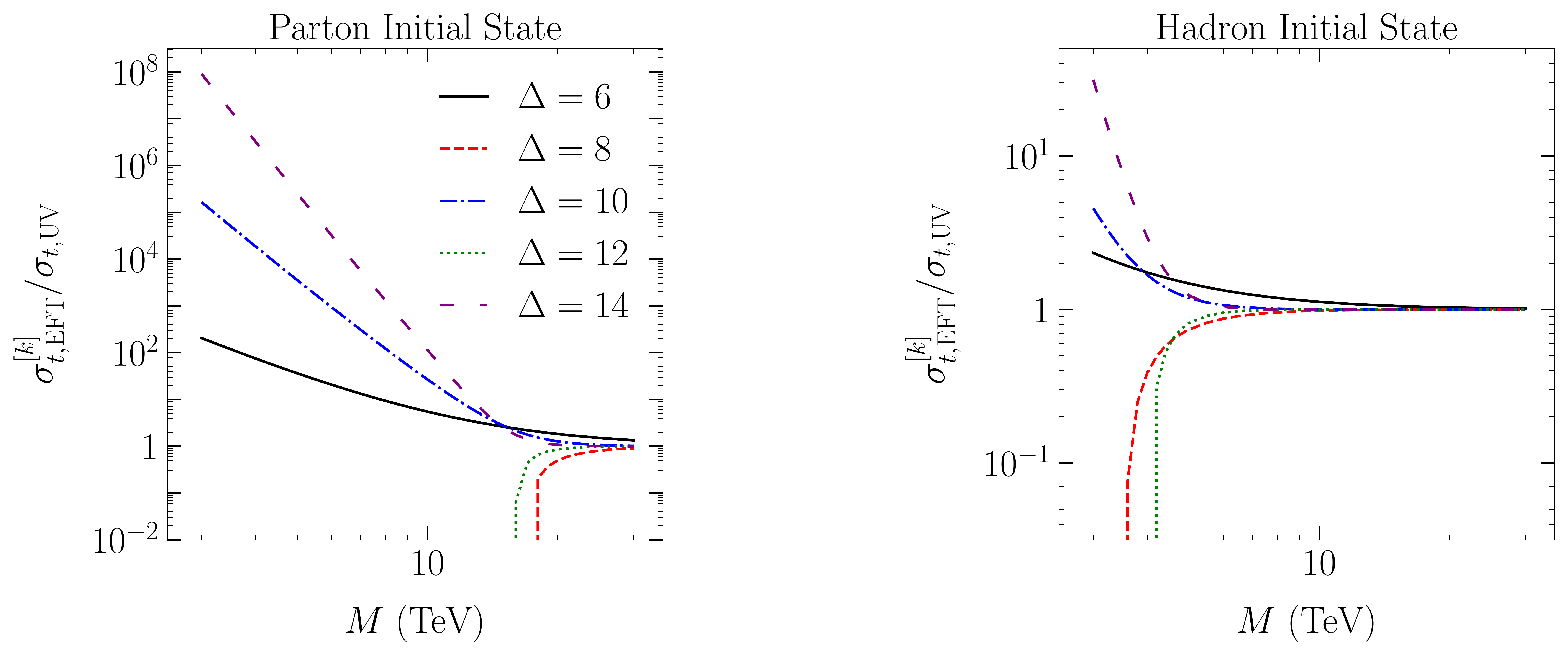}
\caption{The ratio $\sigma_{t,\,\text{EFT}}^{[k]}/\sigma_{t,\s\text{UV}}$ for the first few $\Delta=6+2k$ as a function of $M$.  For the ``Partonic Initial State'' case [left], the series converges for $M>\sqrt{\shat}$ and diverges for $M<\sqrt{\shat}$.  For the ``Hadronic Initial State'' case [right], the series converges for $M>\sqrt{s}$ and appears to be converging when $M\lesssim \sqrt{s}$ (although it actually diverges for $\Delta > \D$).}
\label{fig:sigmatMplot}
\end{figure}

We emphasize that this behavior of the EFT expansion only happens for the parameter space where $M<\sqrt{s}$. If instead the new physics scale $M$ is above the collider energy $\sqrt{s}$, the EFT expansion will yield a convergent series as expected.  The contrast between these two regimes can be seen in \cref{fig:sigmatMplot}, where we plot $\sigma^{[k]}_{t,\s\text{EFT}}/\sigma_{t,\,\text{UV}}$ as a function of $M$ for a few low lying choices of $\Delta$.

\begin{figure}[t]
\centering
\includegraphics[width=0.9\textwidth]{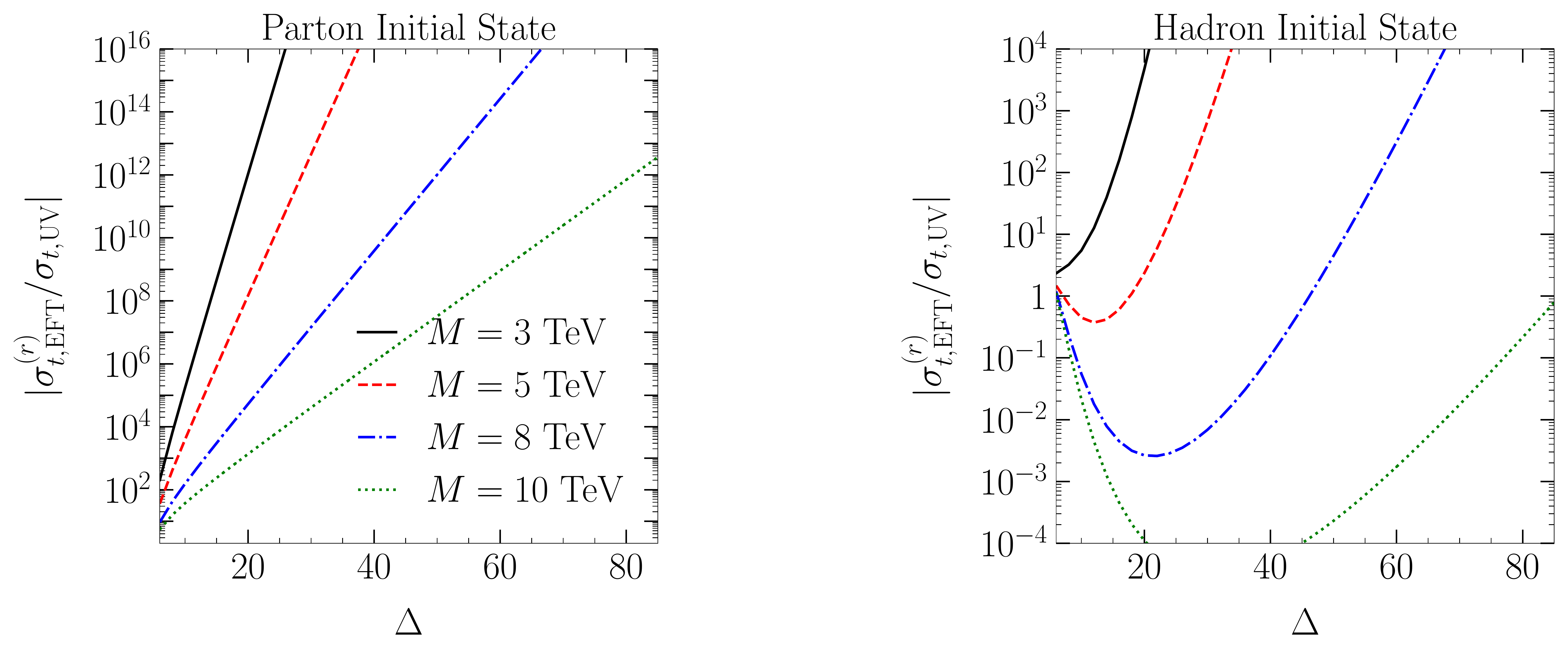}
\caption{The size of the $r^\text{th}$ term $\big| \sigma^{(r)}_{t,\,\text{EFT}}/\sigma_{t,\s\text{UV}} \big|$ as a function of $r$. In the ``Partonic Initial State'' case [left], we have $M<\sqrt{\shat}=14$ TeV; the term grows monotonically with $r$. In the ``Hadronic Initial State'' case [right], we have $M<\sqrt{s}=14$ TeV; the term tends to decrease with $r$ for small $r$ (for $M \gtrsim 5 \text{ TeV}$), and then begins to increase for large $r$.}\label{fig:sigmatAbsTermCollect}
\end{figure}

We can explore the nature of this critical point in the EFT expansion series by investigating the size of its $r^\text{th}$ term $| \sigma_{t,\s\text{EFT}}^{(r)}/\sigma_{t,\,\text{UV}} |$ as a function of $r$, see~\cref{fig:sigmatAbsTermCollect}.
We see that in the parton case, the terms grow monotonically with $r$. This is expected because higher-order terms in the EFT expansion come with more powers of $\shat/M^2>1$. Moving to the hadron case, we see that the terms tend to decrease with $r$ at small $r$ (as long as $M$ is not too small), making the series appear to be converging. This happens because the PDF suppression of high-energy partons brings down the average parton-level center-of-mass energy
\begin{equation}
{\shat}_\text{ave} \equiv \Big(\big \langle\s {\shat}^{\s r}\s \big \rangle_\text{PDF} \Big)^{1/r} \,,
\label{eqn:shateffdef}
\end{equation}
below $M^2$, yielding a suppression factor $\shat_\text{ave}/M^2<1$. However, the size of ${\shat}_\text{ave} \in [0, s]$ of course depends on $r$. As one increases $r$, the effects of $(\shat/M^2)^r$ will eventually win over the PDF suppression factor, causing $\shat_\text{ave}/M^2>1$, which corresponds to where the curves turn around in \cref{fig:sigmatAbsTermCollect}. In fact, ${\shat}_\text{ave}$ becomes infinitely close to the collider energy $s$ as we take $r\to\infty$, so the relative error will always diverge when $s/M^2>1$.

\begin{figure}[t]
\centering
\includegraphics[width=0.4\textwidth]{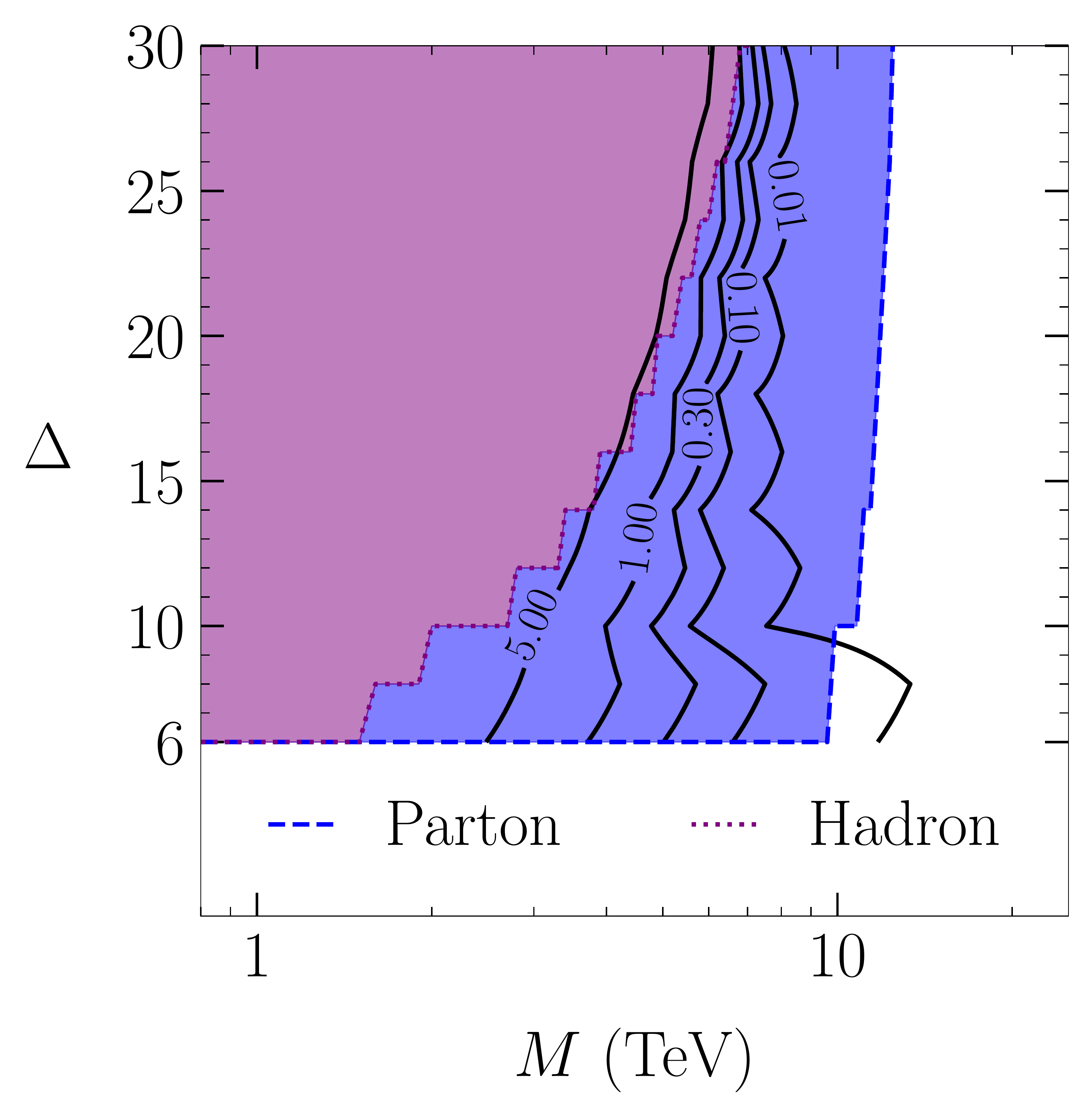}\hspace{35pt}
\includegraphics[width=0.4\textwidth]{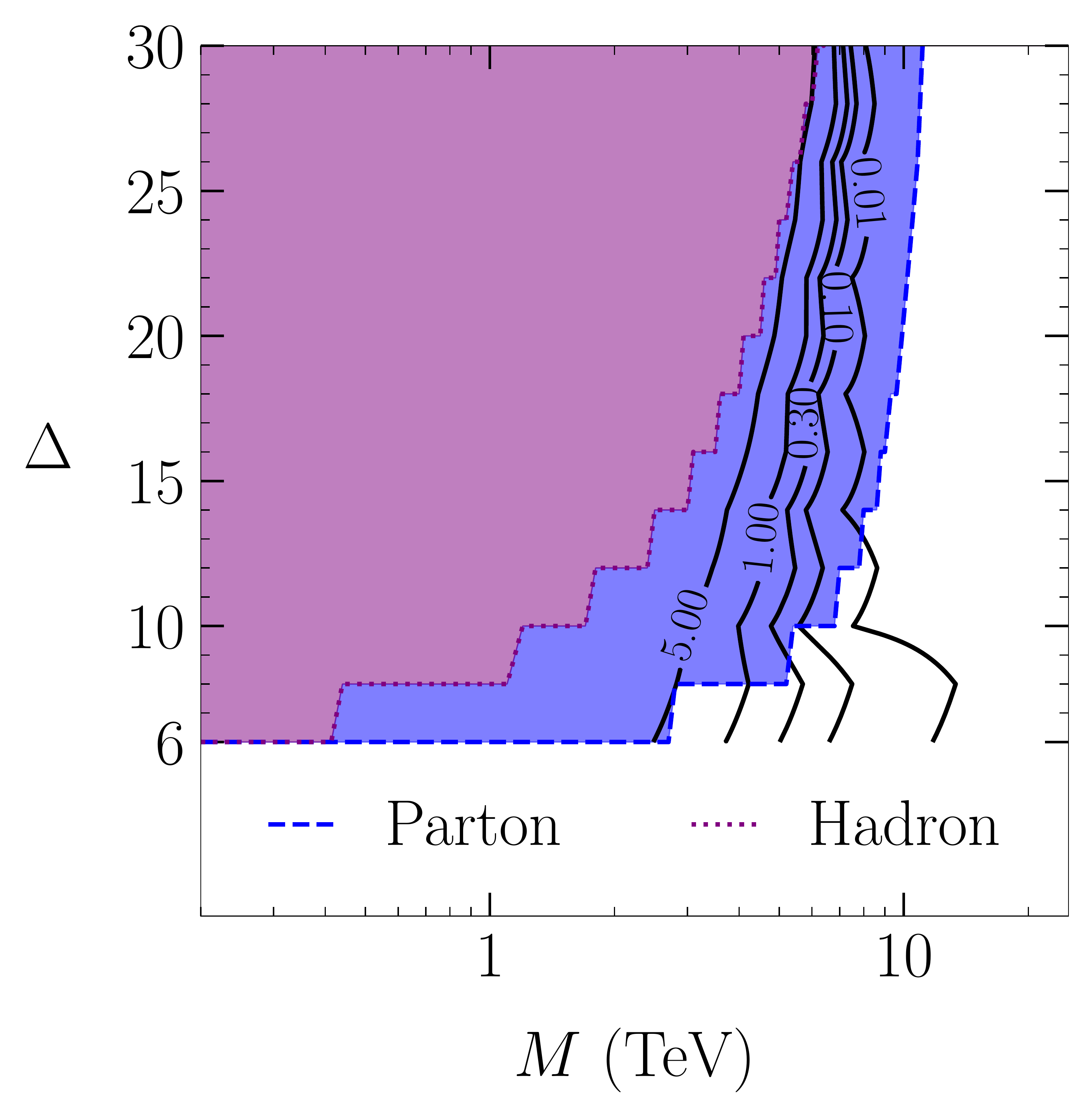}
\caption{A comparison of the perturbative unitarity results against the $t$-channel cross section predictions for two choices of the UV parameters: $\lambda_{q\phi} = 8\pi$ [left] and $\lambda_{q\phi} = 2$ [right].  The shaded regions are the perturbative unitarity bounds. The contours show constant power counting uncertainty. This provides evidence that valid EFTs exist in the region excluded by the naive partonic perturbative unitarity bound.}
\label{fig:OmegatEFTErrorContours}
\end{figure}

Having understood the behavior of the cross section as we vary $M$ and $\Delta$, we can use these results to understand the meaning of the perturbative unitarity bounds derived in the previous section.  In \cref{fig:OmegatEFTErrorContours}, we plot the perturbative unitarity constraint for two points in parameter space, $\lambda_{q\phi} =$ ($8\pi$, $2$) in the (left, right) panel.  Additionally, we overlay contours of constant EFT power counting uncertainty as defined in \cref{eqn:Uncertainty}.\footnote{Due to the alternating behavior in \cref{eqn:hatsigmatEFTr}, the summation to order $k$ is performed separately for even- and odd-valued $r$ terms. We then obtain separate contours for the even and odd sets and interleave them together to restore proper dimension ordering, avoiding a distortion in the contours otherwise.}

As a rough guide, we say that the EFT is providing a good approximation of the underlying UV physics when this uncertainty is $<O(1)$.  We see that when the coupling is large,\footnote{Recall that $\lambda_{q\phi} = 8\pi$ saturates the parton level perturbative unitarity bound.} violating the hadronic perturbative unitarity constraint essentially rules out the region with $> O(1)$ uncertainty.  When the coupling is smaller, there is a region with uncertainty $> O(1)$ that is not excluded by the hadronic perturbative unitarity bound.  This is not a contradiction, since the perturbative unitarity test is only a necessary (but not sufficient) constraint on the validity of the EFT.

We conclude that there are valid EFTs that lie in the region that would be naively excluded by the partonic perturbative unitarity constraint.  Furthermore, the region excluded by considerations of hadronic perturbative unitarity violation do not contain any valid EFTs.

\clearpage
\subsubsection{$s$-channel Results}

\begin{figure}[h!]
\centering
\includegraphics[width=1.0\textwidth]{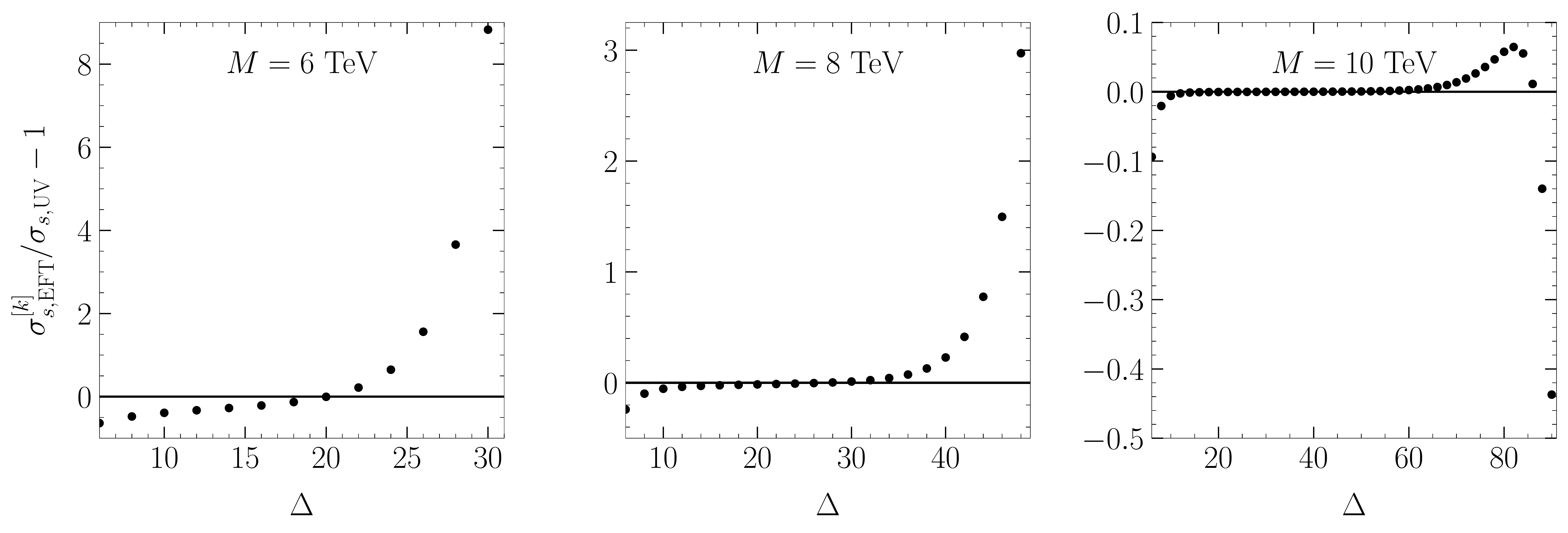}
\caption{The relative error (see \cref{eqn:Errors}) as computed for the $s$-channel model in the ``Hadronic Initial State'' case for $M<\sqrt{s}=14$ TeV and $\BSMphimass=1$ TeV.}\label{fig:sigmasErrorCollect}
\end{figure}

\begin{figure}[h!]
\centering
\includegraphics[width=1.0\textwidth]{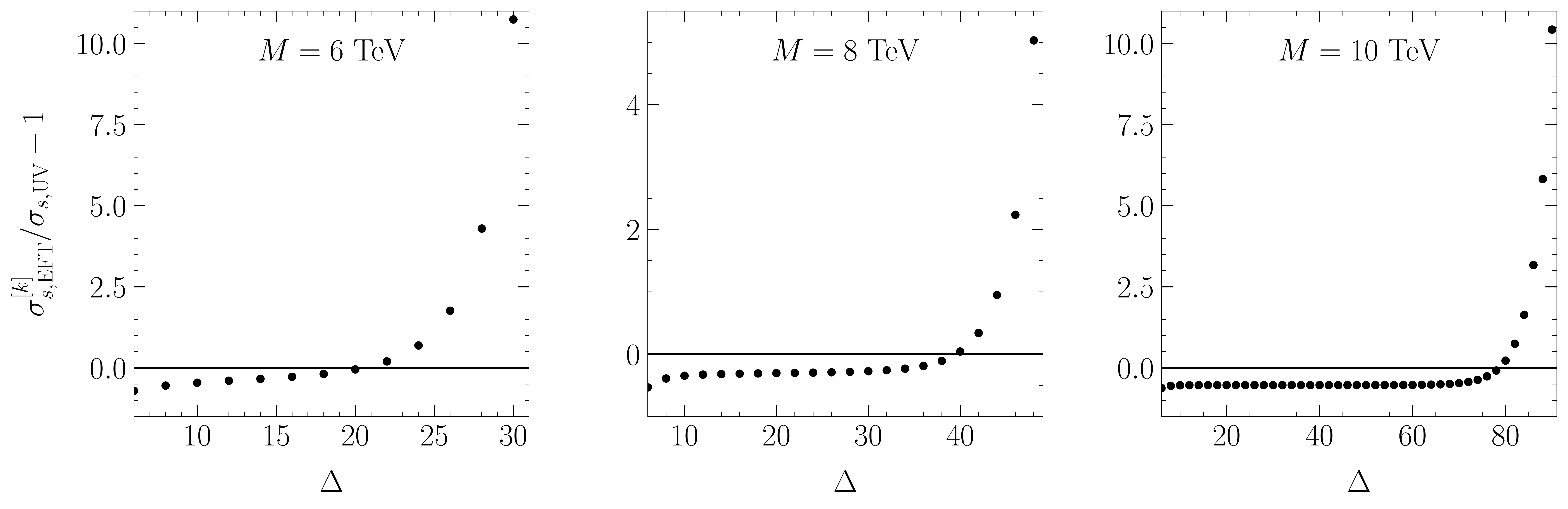}
\caption{The relative error (see \cref{eqn:Errors}) as computed for the $s$-channel model in the ``Hadronic Initial State'' case where $\Gamma \to 0$ for $M<\sqrt{s}=14$ TeV and $\BSMphimass=1$ TeV.  For $\Delta < \D$, the EFT prediction appears to be converging to the wrong value.}\label{fig:sigmasErrorCollect_NoWidth}
\end{figure}

The $s$-channel production results share the same qualitative features as in the $t$-channel case. As before, we study
the relative error defined in \cref{eqn:Errors} as a function of the truncation order $k$ for the case of the $s$-channel model. Typical curves of its absolute value are qualitatively similar to those for the $t$-channel in \cref{fig:sigmatAbsErrorCollect}, where the hadronic EFT expansion also exhibits an apparently converging behavior for small $k$. This feature is elucidated in \cref{fig:sigmasErrorCollect}, where we plot the relative error without taking the absolute value. Note that it is important to include the width effects in the EFT description.  In \cref{fig:sigmasErrorCollect_NoWidth}, we show that taking $\Gamma \to 0$ causes the EFT to converge to the wrong prediction for small $k$. Just as above, the apparently (but actually not) converging behavior of the EFT expansion only happens when the new physics scale $M$ is below the collider energy $\sqrt{s}$; otherwise, the EFT expansion yields a convergent series. The underlying reason for this apparent convergence at small $k$ is again due to the PDF suppression of high-energy partons, which has a non-trivial impact on the relative size between the adjacent terms in the EFT expansion.

\begin{figure}[t]
\centering
\includegraphics[width=0.4\textwidth]{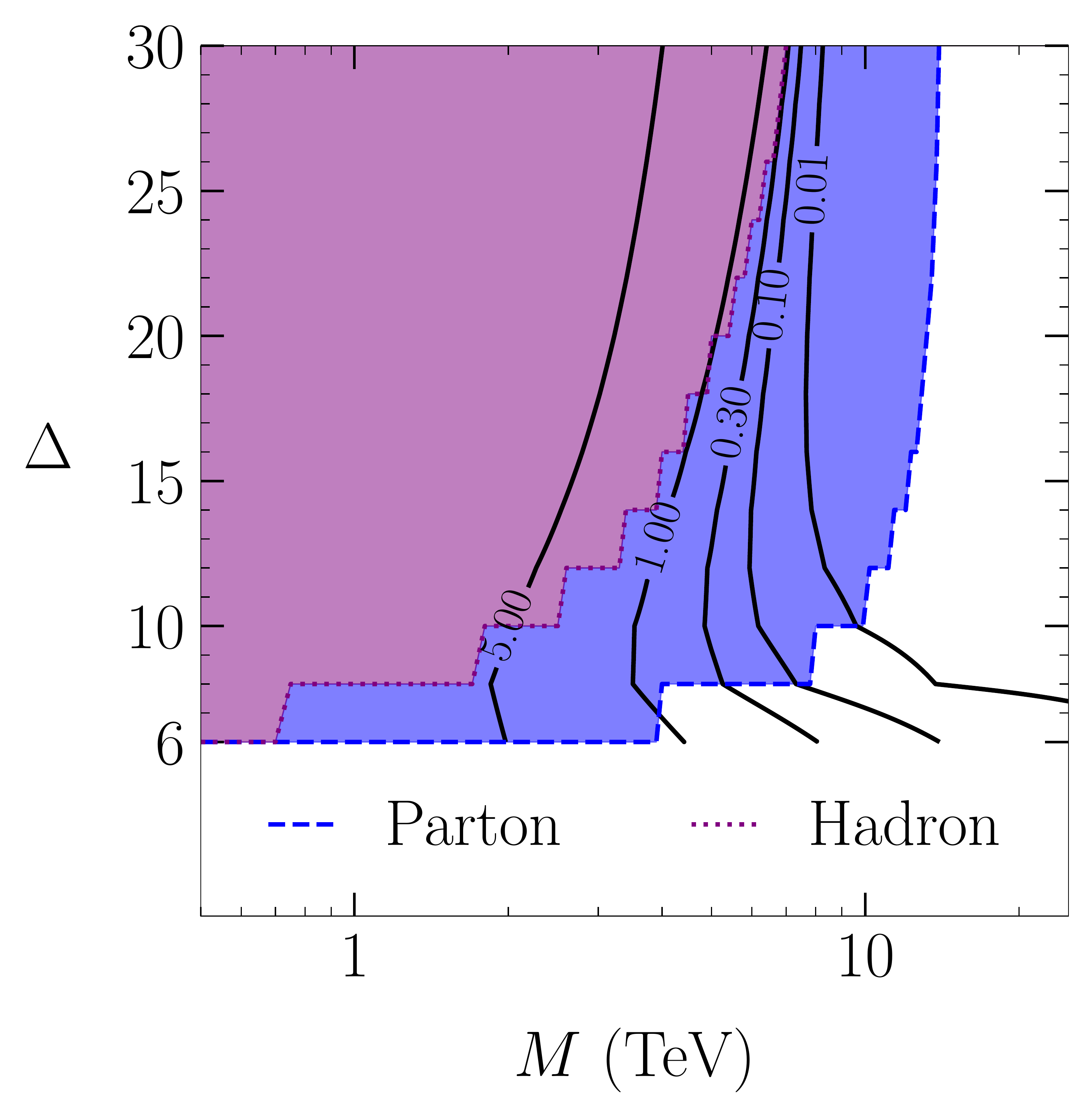}\hspace{35pt}
\includegraphics[width=0.4\textwidth]{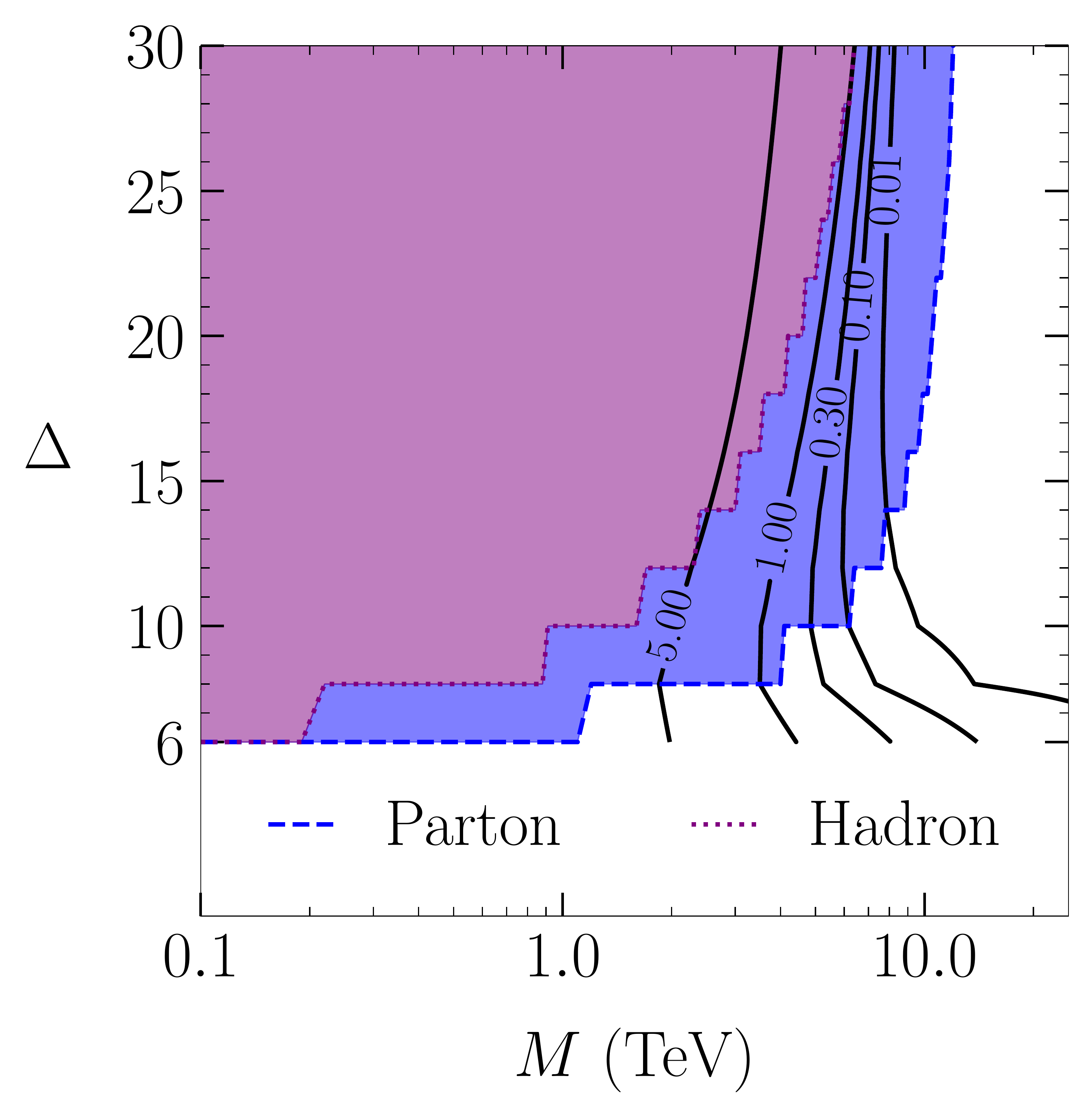}
\caption{A comparison of the perturbative unitarity results against the $s$-channel cross section predictions for two choices of the UV parameters: $\lambda_q = 2$ [left] and $\lambda_q = 2/(4\pi)$ [right].  The shaded regions are the perturbative unitarity bounds. The contours show constant power counting uncertainty. This provides evidence that valid EFTs exist in the region excluded by the naive partonic perturbative unitarity bound.}\label{fig:OmegasEFTErrorContours}
\end{figure}

We plot the perturbative unitarity constraint for two points in parameter space, $\lambda_q =$ ($2$, $2/(4\pi)$) in the (left, right) panel in \cref{fig:OmegasEFTErrorContours}, overlaying contours of constant EFT power counting uncertainty (taking $\Gamma = 0$), as defined in \cref{eqn:Uncertainty}.  Again, this provides evidence for our interpretation that incorporating PDFs into the perturbative unitarity bound is consistent.

\clearpage
\clearpage
\section{Impact of Kinematic Cuts}
\label{sec:Energycuts}

In the previous sections, we explored the extent to which PDFs could soften the high energy contributions enough to maintain the validity of an EFT description.  We saw it is possible that the EFT provides a useful approximation of the full theory even when $M < \sqrt{s}$, as long as one did not include operators with dimension above some critical value.  Our conclusions stem from the essential fact that particle physics scattering is inherently probabilistic, so one needs to collect many events to populate a signal region in order to infer detailed properties of the underlying theory.  In particular, with no further knowledge on the parton-level center-of-mass energy assumed, we computed $\sigma$ and $\Omega$ by integrating over the full kinematic range $\tau \in (\tau_\phi,1)$ (see \cref{eqn:HadronicUnitarityBound,eq:sigmapptopsipsi}), which led us to the quantitative results in the previous sections.

The goal of this section is to explore how sensitive our conclusions are to incorporating additional information about the parton-level kinematics.  Since we are still working in the context of toy models and are only considering 2-to-2 scattering, we will simply focus on just two types of kinematic cuts on the $\tau$ integration range:
\begin{itemize}
\setlength\itemsep{1pt}
\item Cutting away low energy events $\big(\text{requiring } \sqrt{\hat s} > E _{\min} \big)$:~this is a proxy for a set of preselection cuts, including a trigger threshold and/or a minimum cut on a kinematic quantity such as $p_T$, $H_T$, missing energy, \emph{etc}.
\item Cutting away high energy events $\big(\text{requiring } \sqrt{\hat s} < E_{\max}\big)$:~this is a proxy for comparing with a test of EFT validity that is sometimes employed when doing analyses in the parameter space where $M < \sqrt{s}$.  Specifically, we are referring to the test that introduces a cutoff on high energy events and checks that the results are insensitive to this cutoff.
\end{itemize}

\begin{figure}[h]
\centering
\includegraphics[width=0.4\textwidth]{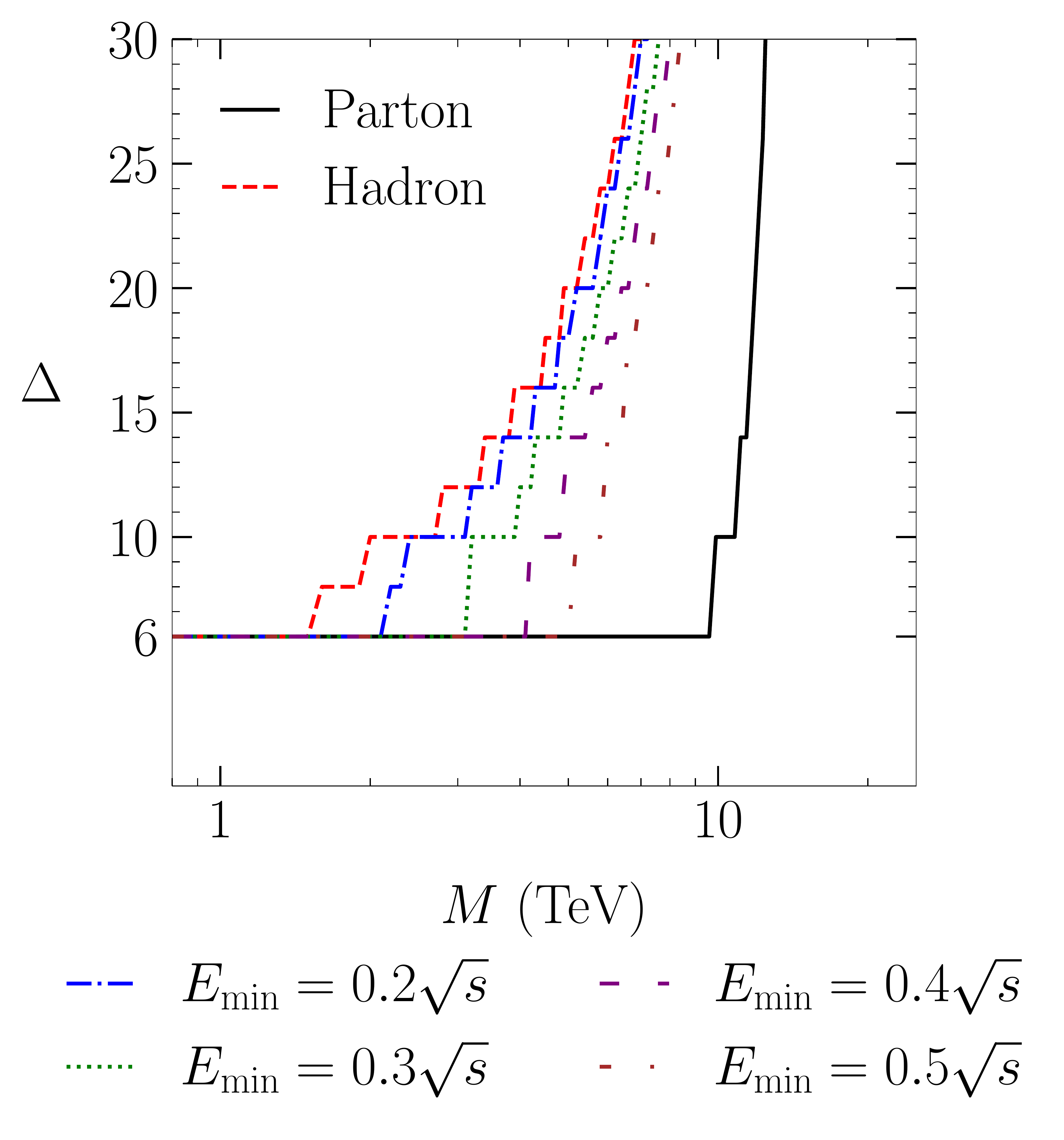}\hspace{35pt}
\includegraphics[width=0.4\textwidth]{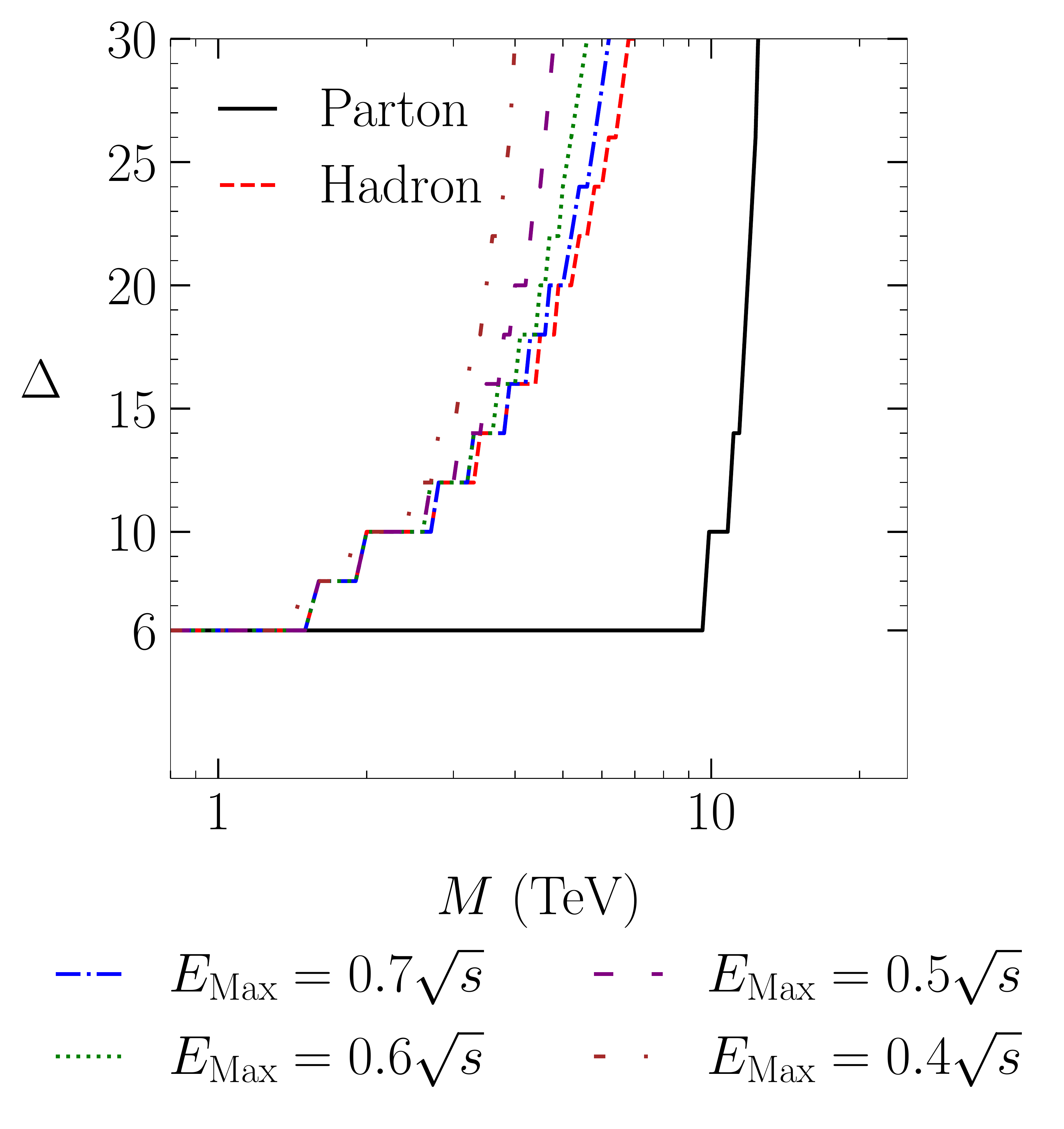}
\caption{Perturbative unitarity bounds in the $\Delta$ versus $M$ plane for various choices of a minimum energy cut $E_{\min}$ [left] and of a maximum energy cut $E_{\max}$ [right] for the $t$-channel model with $\lambda_{q\phi}=8\pi$.  The region that is incompatible with hadronic perturbative partial-wave unitarity is to the left of the curves.}
\label{fig:EVary}
\end{figure}

The results of the study where we vary $E_{\min}$ are presented in the left panel of \cref{fig:EVary}.  We adjust the $E_{\min}$ cut from $0$, labeled ``Hadron'' in the figure, to $0.5 \sqrt{s}$.  We see that the perturbative unitarity bound is not particularly sensitive to the $E_{\min} = 0.2 \sqrt{s}$ cut, but then begins to become stronger quickly. When no cut is applied, the perturbative unitarity bound is roughly $M_\text{bound} \sim 1.5 \text{ TeV}$, as compared to $M_\text{bound} \sim 5 \text{ TeV}$ when the cut is increased to $E_{\min} = 0.5 \sqrt{s}$.  This is a consequence of the shape of the PDFs, which decrease by orders of magnitude as $x$ is increased.  As we emphasized above, the PDFs suppress high energy events, and by increasing the cut on $E_{\min}$ we are essentially removing that suppression which causes the perturbative unitarity bounds to asymptote to the parton result $M_\text{bound} \sim \sqrt{s}$ for $E_{\min} \to \sqrt{s}$ as they should.  When computing the perturbative unitarity bounds on a scenario of interest, it is paramount that these low energy cuts are implemented, since the relevant signal regions often lie in the tails of kinematic distributions (see \eg\ \cite{Farina:2016rws, Lang:2021hnd}).

Next, we turn to the results where we vary $E_{\max}$ presented in the right panel of \cref{fig:EVary}.  This is a proxy for an EFT test that is sometimes utilized, where robustness of the result of an analysis is tested against varying a cutoff on high energy events (see \eg\/ \cite{ATLAS:2014kci, ATLAS:2015qlt, ATLAS:2015wpv, ATLAS:2015ptl, CMS:2015rjz, CMS:2016gox, ATLAS:2016zxj, ATLAS:2017nga, Abercrombie:2015wmb}).  In this scheme, the validity of the result depends on how much the derived limits change as a function of $E_{\max}$.  Typically, even in the parameter space where $M < \sqrt{s}$, results are shown to be relatively insensitive to such a cut.  We can mimic this test by checking that the hadronic perturbative unitarity bound introduced here is robust to varying $E_{\max}$. Indeed, when taking the relatively extreme cut $E_{\max} = 0.4 \sqrt{s}$, the bounds on $M$ barely change for EFTs with relatively low truncation dimensions, which is relevant for most of practical applications. This confirms that our bounds are compatible with this test. On the other hand, for large $\Delta$, $M_\text{bound}$ gets weaker with the $E_{\max}$ cut, consistent with our expectation that as we take $\Delta$ to be large, $M_\text{bound} \rightarrow E_{\max}$.

\clearpage

\section{Discussion and Future Directions}
\label{sec:Discussions}

In this paper, we have studied the validity of the EFT framework as it applies to searching for new physics associated with a scale that is below the center-of-mass energy at hadron colliders $M < \sqrt{s}$. The key insight is that when the signal regions are designed to be inclusive regarding the partonic center-of-mass energy, one needs to carefully account for PDF effects, which serve to suppress events that have a high partonic center-of-mass energy.  Using the tree-level pair production process in~\cref{eqn:pairproduction} as a benchmark, we have probed this question in the context of perturbative partial-wave unitarity constraints.  We conclude that there exists parameter space where the EFT defined with $\Delta < \D$ does not violate partial wave perturbative unitarity, even though $M\ll \sqrt{s}$. We provided evidence that there exist valid EFTs that lie in the parameter space opened up by PDF effects.  Importantly, this conclusion is of practical relevance to EFT analyses being performed at the LHC, which often result in limits on the EFT scale that are below $\sqrt{s}$.  We emphasize that the perturbative unitarity constraint depends on the kinematics of the process being studied, so for each given search one must perform a dedicated analysis to obtain the EFT parameter space compatible with perturbative partial-wave unitarity.

We view this paper as demonstrating that PDF suppression of high energy events can dramatically increase the valid region of parameter space for an EFT search.  The most obvious next step is to compute the perturbative unitarity bounds of realistic EFT extensions of the Standard Model, \eg\ of relevance to dark matter searches, for constraining SMEFT operator coefficients, \emph{etc}.  Specifically, we would like to revisit the perturbative partial-wave unitarity bound to incorporate fermionic initial states, and then to apply this upgraded calculation to specific EFTs that are being searched for at the LHC.  As we discussed in \cref{sec:Energycuts}, the details of the signal region cuts will also have an impact on the detailed bounds.  This analysis will be critical to applying our results in detail at the LHC.  It would additionally be interesting to understand the interplay between PDF fits and the inclusion of higher dimension SMEFT operators, along the lines of~\cite{Madigan:2021uho}.

We also plan to investigate how our findings here generalize beyond the specific 2-to-2 process of~\cref{eqn:pairproduction}. In particular, \cref{eqn:lagsEFT,eqn:lagtEFT} show that the EFT expansions for this process are only accounting for higher-dimension operators that strictly involve more powers of derivatives. In general, one would like to see that the same conclusions hold when including operators that involve more powers of fields. While we anticipate that our conclusions will be essentially unchanged when studying the effects of these operators due simply to dimensional analysis presented in \cref{subsec:PowerCounting}, it will be interesting to see how the interplay of making inclusive/exclusive requirements on the final states will impact the bounds derived here.  It will additionally be useful to explore EFT validity in the context of experimental limits that are placed with shape information.

Even if we restrict our scope to the derivative expansion as was done in this paper, there are potential applications that could follow up on some recent studies where resumming the EFT field expansion was utilized:
\begin{itemize}
\setlength\itemsep{1pt}
\item Refs.~\cite{Alonso:2015fsp, Alonso:2016oah, Cohen:2020xca, Banta:2021dek} argued that one must include all orders in the field expansion in order to correctly identify if a BSM EFT can be matched onto SMEFT (as opposed to being forced to match onto the more general formulation with non-linearly realized electroweak symmetry breaking).  There is additionally a close relation between perturbative unitarity violation and inclusive amplitudes involving an arbitrary number of fields in the final state~\cite{Falkowski:2019tft, Cohen:2021ucp}.
\item When focusing on BSM modifications to the two-point and three-point amplitudes, it was emphasized that the derivative expansion is trivial \cite{Henning:2017fpj, Helset:2020yio}. In these cases, one can resum the field expansion in SMEFT, an approach that was recently advocated in~\cite{Helset:2020yio}.
\item Going beyond two-point and three-point amplitudes, a resummation over the field expansion will leave us with a non-trivial derivative expansion \cite{Helset:2020yio}. Nevertheless, the derivative expansion can still be systematically organized through the use of group theoretical techniques~\cite{Elvang:2020lue,Henning:2017fpj}.
\end{itemize}
Once the field expansion is resummed, the EFT will only include a derivative expansion. Therefore, our study here helps to justify the validity of analyses that resum (some of) the field expansion, and it would be of significant interest to understand the interplay between these ideas and PDF effects.

Finally, we will briefly comment on the implications for new search designs at the LHC. Given the dependence of the perturbative unitarity bound on the kinematics used to define the signal region of the search, one could be motivated to narrow the range of final state energies to sharpen the perturbative unitarity bound. However, this  typically increases the statistical error, thereby reducing the power of the search. Furthermore, since the meaning of the unitarity bound is limited to the assumption of perturbativity, it is unclear how much is gained by attempting to sharpen it by modifying the search strategy. Based on these considerations, we believe that in many cases, it might be advantageous to keep the energy bin of the experimental search somewhat inclusive. Investigating this interplay is worthy of dedicated studies, which we leave for future work.

Analyses that utilize EFTs are of critical importance to the LHC program.  Accounting for the impact of PDFs on their range of validity will allow us to utilize these frameworks with confidence as we continue to pursue the experimental signatures of beyond the Standard Model physics.

\acknowledgments
%
T.C.\ is especially grateful to Nima Arkani-Hamed for the conversation that inspired this work.
We thank Spencer Chang, Matt Dolan, Graham Kribs, Markus Luty, Adam Martin, Francesco Riva, and David Strom for valuable conversations and feedback on the manuscript.
T.C.,\ J.D.,\ and X.L.\ are supported by the U.S. Department of Energy, under grant number DE-SC0011640.
%

\appendix
\section*{Appendix}
\addcontentsline{toc}{section}{\protect\numberline{}Appendix}%
\renewcommand*{\thesubsection}{\Alph{subsection}}
\numberwithin{equation}{subsection}
\section{Results for Smaller Final State Mass}
\label{appsec:BSMpsimassLow}

\begin{figure}[h]
\centering
\includegraphics[width=0.4\textwidth]{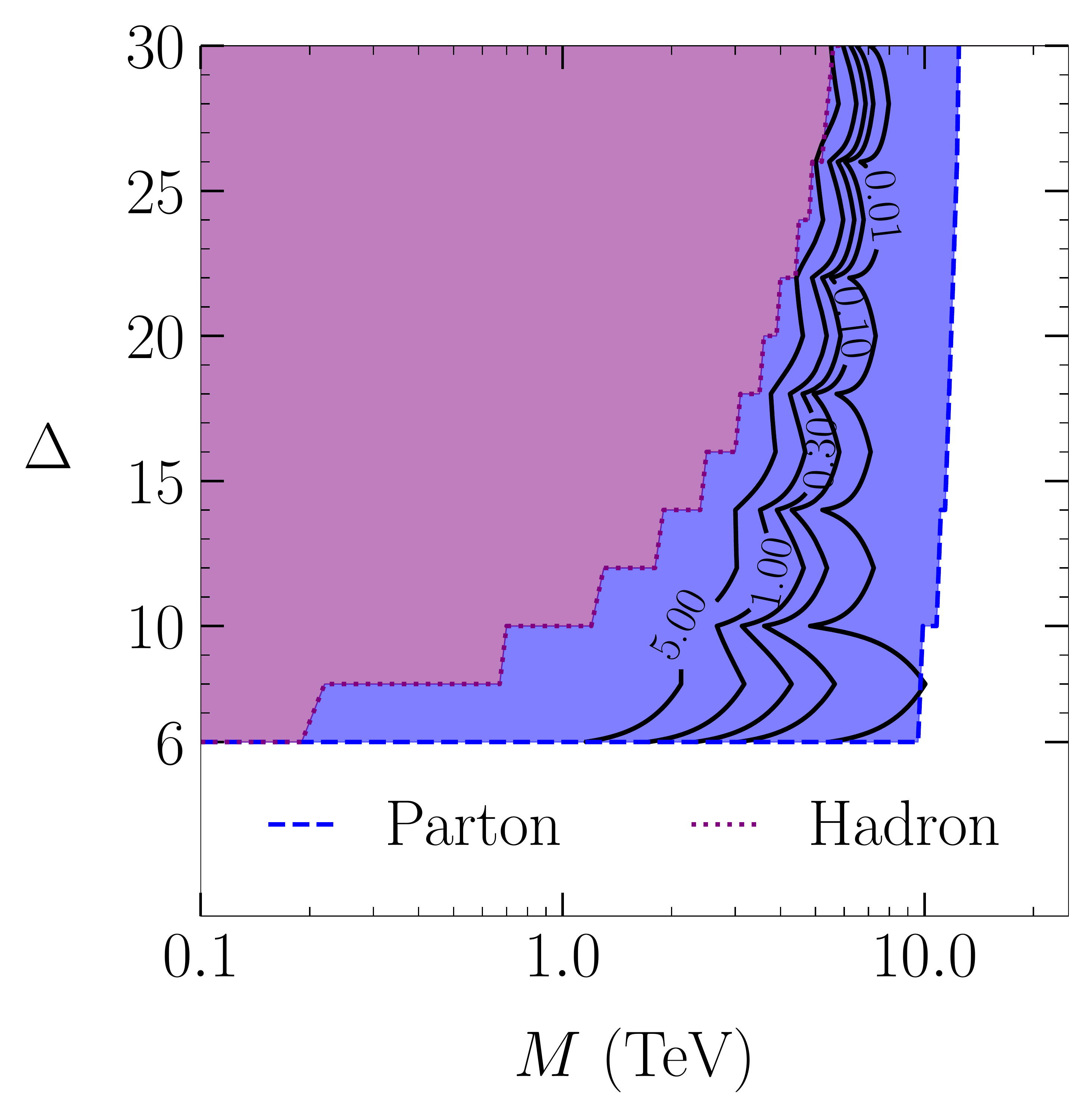}\hspace{35pt}
\includegraphics[width=0.4\textwidth]{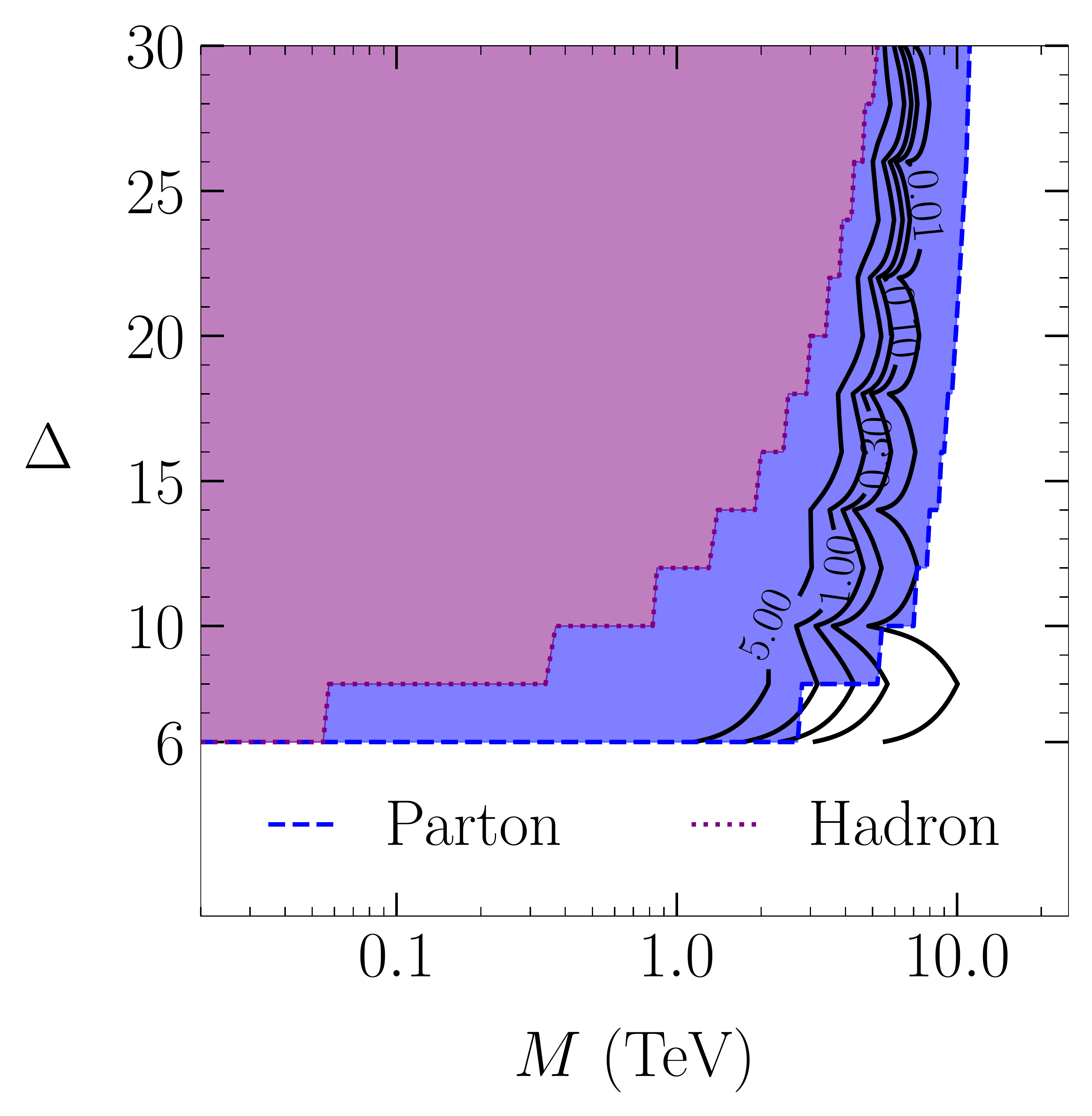}
\caption{A comparison of the perturbative unitarity results against the $t$-channel cross section predictions for two choices of the UV parameters: $\lambda_{q\phi} = 8\pi$ [left] and $\lambda_{q\phi} = 2$ [right] in the case that the final state particles have a mass of 10 GeV.  The shaded regions are the perturbative unitarity bounds, while the contours show power counting error.}
\label{fig:OmegatEFTErrorContoursLight}
\end{figure}

\begin{figure}[h]
\centering
\includegraphics[width=0.4\textwidth]{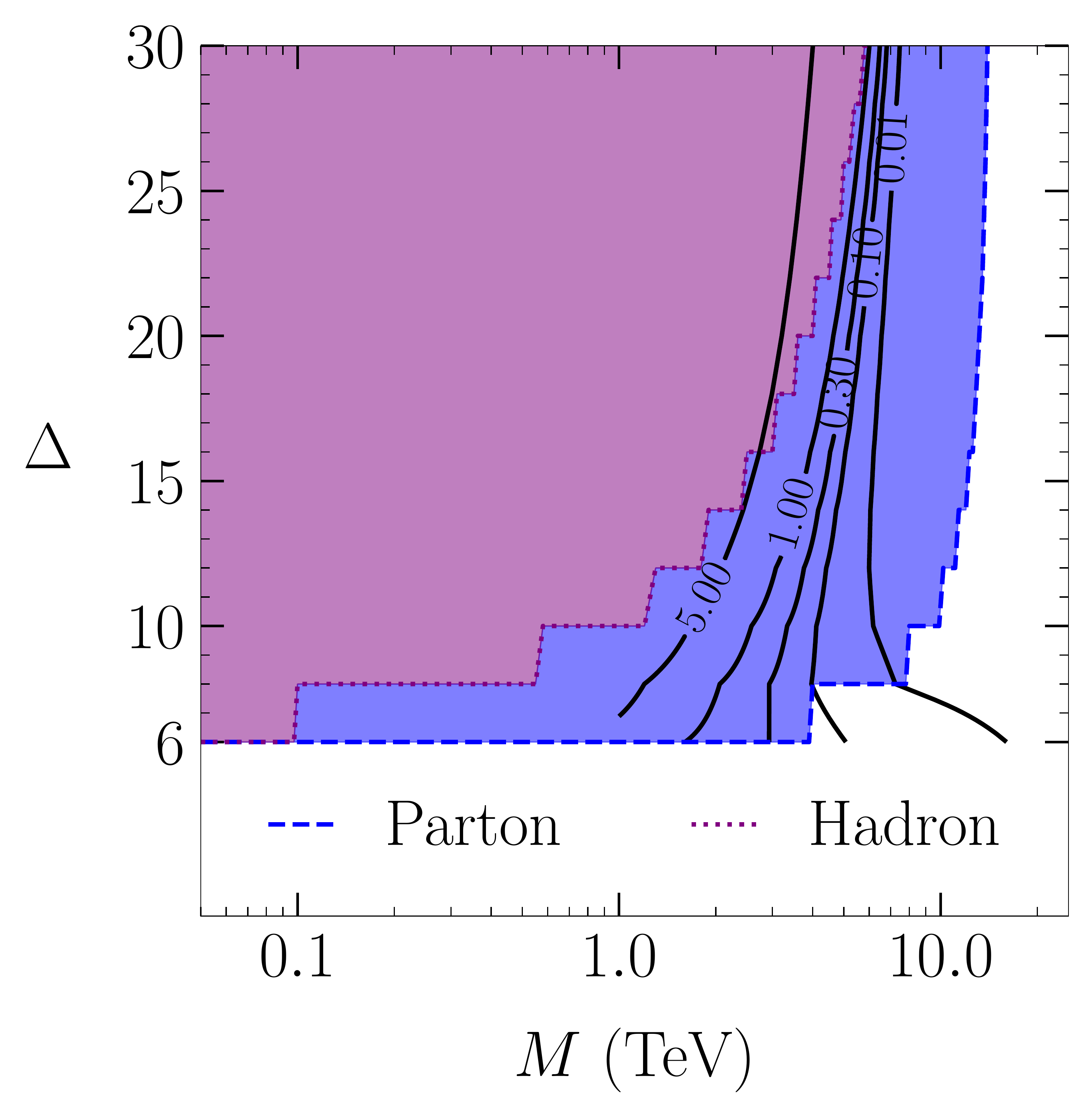}\hspace{35pt}
\includegraphics[width=0.4\textwidth]{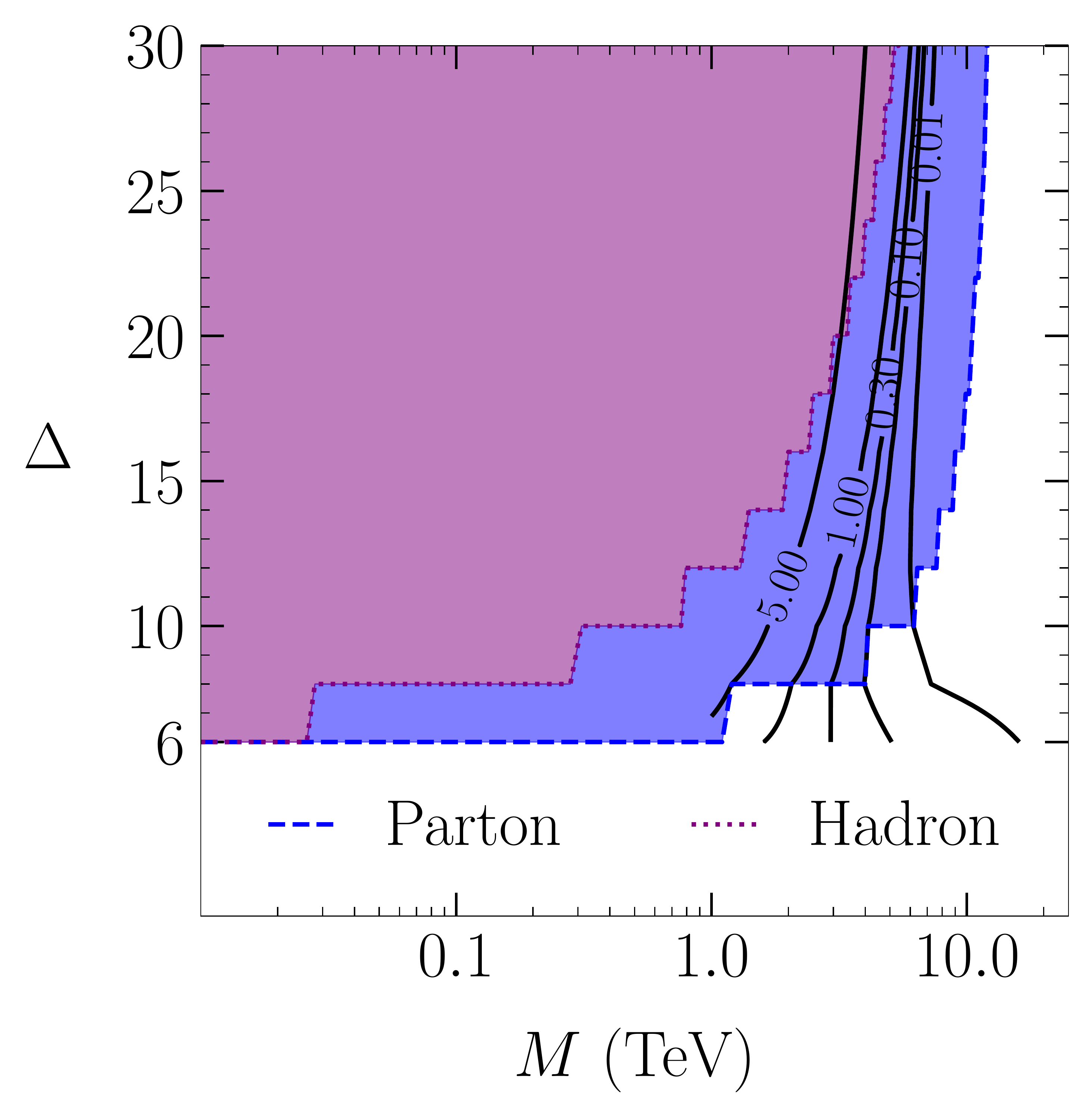}
\caption{A comparison of the perturbative unitarity results against the $s$-channel cross section predictions for two choices of the UV parameters: $\lambda_q = 2$ [left] and $\lambda_q = 2/(4\pi)$ [right] in the case that the final state particles have a mass of 10 GeV. The shaded regions are the perturbative unitarity bounds, while the contours show power counting error.}
\label{fig:OmegasEFTErrorContoursLight}
\end{figure}

\begin{figure}[h]
\centering
\includegraphics[width=0.4\textwidth]{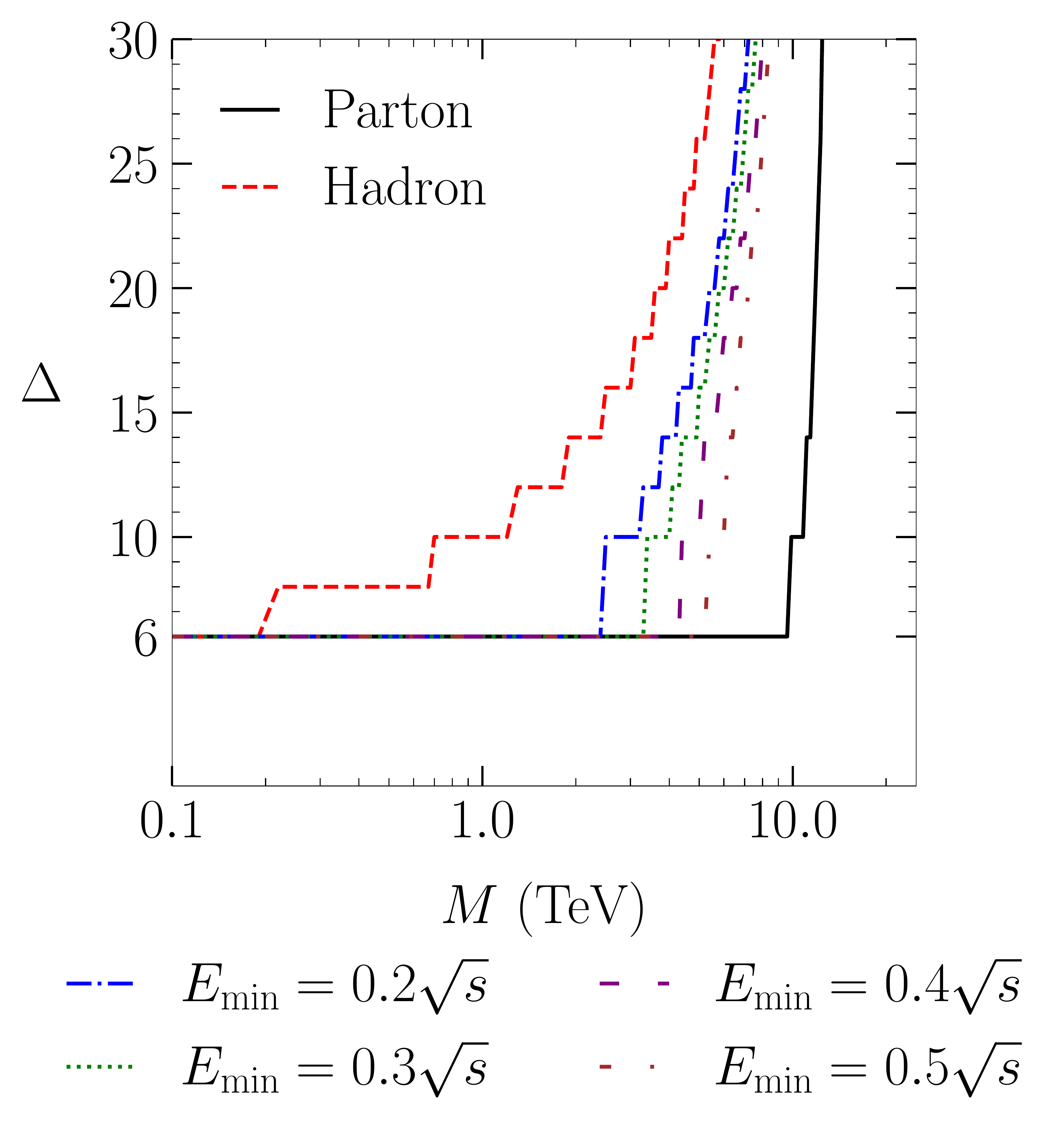}\hspace{35pt}
\includegraphics[width=0.4\textwidth]{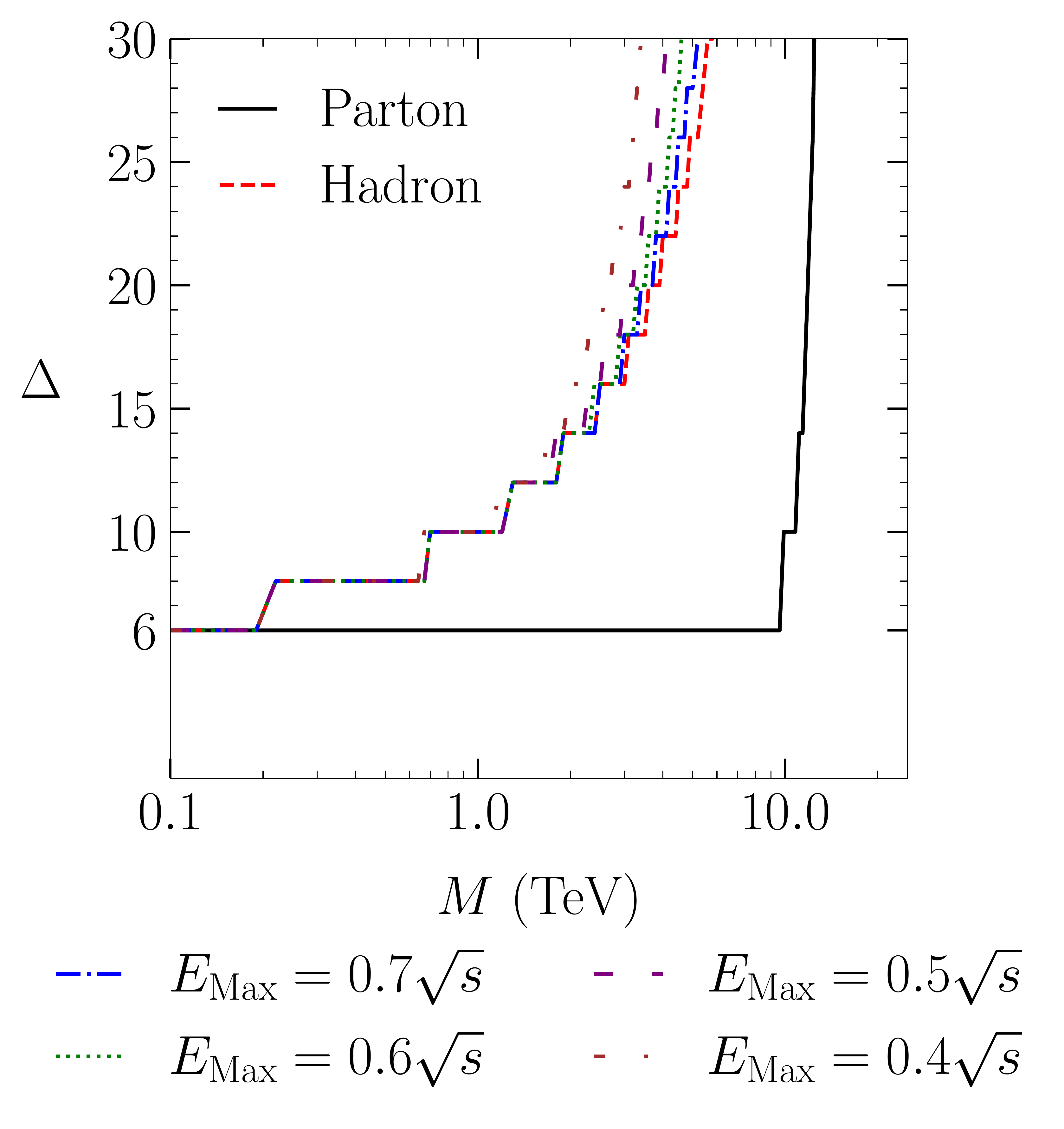}
\caption{Perturbative unitarity bounds in the $\Delta$ versus $M$ plane for various choices of a minimum energy cut $E_{\min}$ [left] and of a maximum energy cut $E_{\max}$ [right] for the $t$-channel model with $\lambda_{q\phi}=8\pi$ in the case that the final state particles have a mass of 10 GeV.  The region that is incompatible with hadronic perturbative partial-wave unitarity is to the left of the curves.}
\label{fig:EVaryLight}
\end{figure}

In this Appendix, we provide the perturbative unitarity bounds on the EFT parameter space $(M, \Delta)$ for the case that the final state particles have a mass of $10 \text{ GeV}$, see \cref{fig:OmegatEFTErrorContoursLight,fig:OmegasEFTErrorContoursLight}. The impact of cutting away low (high) energy events is shown in the left (right) panel of \cref{fig:EVaryLight}.

\clearpage

\end{spacing}

\begin{spacing}{1.09}
\addcontentsline{toc}{section}{\protect\numberline{}References}%
\bibliographystyle{utphys}
\bibliography{EFT}
\end{spacing}
\end{document}